\documentstyle[12pt]{article}
\newlength{\pubnumber} 

\catcode`\@=11
\@addtoreset{equation}{section}

\def\section{\@startsection{section}{1}{\z@}{3.5ex plus 1ex minus .2ex}
 {2.3ex plus .2ex}{\large\bf}}
\def\subsection{\@startsection{subsection}{2}{\z@}{2.3ex plus .2ex}
 {2.3ex plus .2ex}{\bf}}

\begin{document}

\begin{titlepage}
\samepage{
\rightline{CERN-TH/98-259}
\rightline{LPTHE-ORSAY 98/56}
\rightline{\tt hep-ph/9809406}
\rightline{September 1998}
\vfill
\begin{center}
   {\Large \bf Cosmological Phase Transitions and\\
     \smallskip
    Radius Stabilization in Higher Dimensions}
\vfill
\vspace{.08in}
   {\large
    Keith R. Dienes$^1$,$\,$
    E. Dudas$^{1,2}$,$\,$
      T. Gherghetta$^1$,$\,$
       A. Riotto$^1$\footnote{
     E-mail addresses: keith.dienes, emilian.dudas, tony.gherghetta,
          antonio.riotto$\,$@cern.ch.}\footnote{
      On leave of absence from the University of Oxford,
           Theoretical Physics, Oxford, UK. }
    \\}
\vspace{.18in}
 {\it  $^1$ CERN Theory Division, CH-1211 Geneva 23, Switzerland\\}
\vspace{.04in}
 {\it  $^2$ LPTHE, Univ.\ Paris-Sud, F-91405 Orsay Cedex, France\footnote{
         Laboratoire associ\'e au CNRS-URA-D0063.}\\}
\end{center}
\vfill
\begin{abstract}
  {\rm
    Recently there has been considerable interest in field theories
    and string theories with large extra spacetime dimensions.
    In this paper, we explore the role of such extra dimensions for
    cosmology, focusing on cosmological phase transitions in field theory and
    the Hagedorn transition and radius stabilization in string theory.
    In each case, we find that significant distinctions emerge
    from the usual case in which such large extra dimensions are absent.
    For example, for temperatures larger than the scale of the compactification
    radii, we show that the critical temperature above which symmetry
    restoration occurs
    is reduced relative to the usual four-dimensional case,
        and consequently cosmological phase transitions in extra dimensions are
         delayed.   Furthermore, we argue that if phase transitions do occur at
       temperatures larger than the compactification scale, then they cannot
      be of first-order type.
    Extending our analysis to string theories with large internal dimensions,
     we focus on the Hagedorn transition and the new features that
   arise due to the presence of large internal dimensions.
       We also consider the
     role of thermal effects in establishing a potential for the radius
      of the compactified dimension, and we use this to propose a thermal
mechanism
         for generating and stabilizing a large radius of compactification.  }
\end{abstract}
\vspace{.13in}  }
\end{titlepage}

\setcounter{footnote}{0}

\newcommand{\newc}{\newcommand}

\newc{\gsim}{\lower.7ex\hbox{$\;\stackrel{\textstyle>}{\sim}\;$}}
\newc{\lsim}{\lower.7ex\hbox{$\;\stackrel{\textstyle<}{\sim}\;$}}

\def\beq{\begin{equation}}
\def\eeq{\end{equation}}
\def\beqn{\begin{eqnarray}}
\def\eeqn{\end{eqnarray}}
\def\sosixteen{{$SO(16)\times SO(16)$}}
\def\e8{{$E_8\times E_8$}}
\def\V#1{{\bf V_{#1}}}
\def\half{{\textstyle{1\over 2}}}
\def\ttwo{{\vartheta_2}}
\def\tthree{{\vartheta_3}}
\def\tfour{{\vartheta_4}}
\def\ttwob{{\overline{\vartheta}_2}}
\def\tthreeb{{\overline{\vartheta}_3}}
\def\tfourb{{\overline{\vartheta}_4}}
\def\etainv{{\overline{\eta}}}
\def\Str{{{\rm Str}\,}}
\def\bone{{\bf 1}}
\def\chibar{{\overline{\chi}}}
\def\Jbar{{\overline{J}}}
\def\qbar{{\overline{q}}}
\def\calO{{\cal O}}
\def\calE{{\cal E}}
\def\calT{{\cal T}}
\def\calM{{\cal M}}
\def\calF{{\cal F}}
\def\calY{{\cal Y}}
\def\rep#1{{\bf {#1}}}
\def\ie{{\it i.e.}\/}
\def\eg{{\it e.g.}\/}
\def\eleven{{(11)}}
\def\ten{{(10)}}
\def\nine{{(9)}}
\def\Ip{{\rm I'}}
\def\oneprime{{I$'$}}
\hyphenation{su-per-sym-met-ric non-su-per-sym-met-ric}
\hyphenation{space-time-super-sym-met-ric}
\hyphenation{mod-u-lar mod-u-lar--in-var-i-ant}


\def\inbar{\,\vrule height1.5ex width.4pt depth0pt}

\def\IC{\relax\hbox{$\inbar\kern-.3em{\rm C}$}}
\def\IQ{\relax\hbox{$\inbar\kern-.3em{\rm Q}$}}
\def\IR{\relax{\rm I\kern-.18em R}}
 \font\cmss=cmss10 \font\cmsss=cmss10 at 7pt
\def\IZ{\relax\ifmmode\mathchoice
 {\hbox{\cmss Z\kern-.4em Z}}{\hbox{\cmss Z\kern-.4em Z}}
 {\lower.9pt\hbox{\cmsss Z\kern-.4em Z}}
 {\lower1.2pt\hbox{\cmsss Z\kern-.4em Z}}\else{\cmss Z\kern-.4em Z}\fi}

\def\NPB#1#2#3{{\it Nucl.\ Phys.}\/ {\bf B#1} (19#2) #3}
\def\PLB#1#2#3{{\it Phys.\ Lett.}\/ {\bf B#1} (19#2) #3}
\def\PRD#1#2#3{{\it Phys.\ Rev.}\/ {\bf D#1} (19#2) #3}
\def\PRL#1#2#3{{\it Phys.\ Rev.\ Lett.}\/ {\bf #1} (19#2) #3}
\def\PRT#1#2#3{{\it Phys.\ Rep.}\/ {\bf#1} (19#2) #3}
\def\CMP#1#2#3{{\it Commun.\ Math.\ Phys.}\/ {\bf#1} (19#2) #3}
\def\MODA#1#2#3{{\it Mod.\ Phys.\ Lett.}\/ {\bf A#1} (19#2) #3}
\def\IJMP#1#2#3{{\it Int.\ J.\ Mod.\ Phys.}\/ {\bf A#1} (19#2) #3}
\def\NUVC#1#2#3{{\it Nuovo Cimento}\/ {\bf #1A} (#2) #3}
\def\etal{{\it et al.\/}}

\def\ZPC#1#2#3{{\it Z.~Phys.}\/ {\bf C#1} (19#2) #3}
\def\PTP#1#2#3{{\it Prog.~Theor.~Phys.}\/ {\bf#1}  (19#2) #3}
\def\MPLA#1#2#3{{\it Mod.~Phys.~Lett.}\/ {\bf#1} (19#2) #3}
\def\AP#1#2#3{{\it Ann.~Phys.}\/ {\bf#1} (19#2) #3}
\def\RMP#1#2#3{{\it Rev.~Mod.~Phys.}\/ {\bf#1} (19#2) #3}
\def\HPA#1#2#3{{\it Helv.~Phys.~Acta}\/ {\bf#1} (19#2) #3}
\def\JETPL#1#2#3{{\it JETP~Lett.}\/ {\bf#1} (19#2) #3}
\def\JPG#1#2#3{{\it J.~Phys.~G.}\/{\bf G#1}(19#2) #3}

\def\ifmath#1{\relax\ifmmode #1\else $#1$\fi}
\def\br{{\rm BR}}
\def\hl{h^0}
\def\mhl{m_{\hl}}
\def\msTop{m_{\,\widetilde{T}}}
\def\mstop{m_{\,\widetilde{t}}}
\def\mstopeff{m_{\,\widetilde{t}}^{\rm eff}}
\def\mlsq{m\ls{L}^2}
\def\mtsq{m\ls{T}^2}
\def\mdsq{m\ls{D}^2}
\def\mssq{m\ls{S}^2}
\def\calm{{\cal M}}
\def\calv{{\cal V}}
\def\crrr{\cr\noalign{\vskip8pt}}
\def\rta{\rightarrow}
\def\bold#1{\setbox0=\hbox{$#1$}%
     \kern-.025em\copy0\kern-\wd0
     \kern.05em\copy0\kern-\wd0
     \kern-.025em\raise.0433em\box0 }
\def\eighth{\ifmath{{\textstyle{1 \over 8}}}}
\def\GENITEM#1;#2{\par\vskip6pt \hangafter=0 \hangindent=#1
   \Textindent{$ #2$ }\ignorespaces}
\def\normalrefmark#1{$^{\scriptstyle #1}$}
\def\unlock{\catcode`@=11} 
\def\lock{\catcode`@=12} 
\def\esphal{E_{\rm sph}}
\def\be{\begin{equation}}
\def\ee{\end{equation}}
\def\bea{\begin{eqnarray}}
\def\eea{\end{eqnarray}}
\def\simlt{\stackrel{<}{{}_\sim}}
\def\simgt{\stackrel{>}{{}_\sim}}
\def\Im{\mathop{\rm Im}}
\def\ov{\overline}

\def\Re{\mathop{\rm Re}}
\def\Tr{\mathop{\rm Tr}}
\def\und{\underline}
\def\dalpha{{\dot\alpha}}
\def\dbeta{{\dot\beta}}
\def\drho{{\dot\rho}}
\def\dsigma{{\dot\sigma}}
\def\crbig{\\\noalign{\vspace {3mm}}}
\def\bigint{{\displaystyle\int}}
\def\Fcomp{{\theta\theta}}
\def\Fbarcomp{\ov{\theta\theta}}
\def\Dcomp{{\theta\theta\ov{\theta\theta}}}
\def\Dint{{\bigint d^2\theta d^2\ov\theta\,}}
\def\Fint{{\bigint d^2\theta\,}}
\def\Fbarint{{\bigint d^2\ov\theta\,}}
\def\ex{{\rm exp}}
\def\mst1{m_{\;\widetilde{t}_{1}}}
\def\msti{m_{\;\widetilde{t}_i}}
\def\mstj{m_{\;\widetilde{t}_j}}
\def\msbi{m_{\;\widetilde{b}_i}}
\def\msbj{m_{\;\widetilde{b}_j}}
\def\st{\;\widetilde{t}}
\def\sb{\;\widetilde{b}}
\def\v{\varphi_c}
\def\mst2{m_{\;\widetilde{t}_{2}}}
\def\mst12{m_{\;\widetilde{t}_{1,2}}}
\def\mstlr{m_{\;\widetilde{t}_{L,R}}}
\def\mstl{m_{\;\widetilde{t}_L}}
\def\mstr{m_{\;\widetilde{t}_R}}
\def\MSbar{\overline{\rm MS}}
\def\DRbar{\overline{\rm DR}}
\def\msb1{m_{\;\widetilde{b}_{1}}}
\def\msb2{m_{\;\widetilde{b}_{2}}}
\def\msb12{m_{\;\widetilde{b}_{1,2}}}
\def\msbl{m_{\;\widetilde{b}_L}}
\def\msbr{m_{\;\widetilde{b}_R}}
\def\modh{\left|H\right|}
\def\mtilde2{\widetilde{m}^{2}}
\def\lambdatilde{\widetilde{\lambda}}
\def\Lambdatilde{\widetilde{\Lambda}}
\def\leff{\lambda_{\rm eff}}
\def\mbart{\overline{m}_{t}}
\def\exis{\varphi}

\long\def\@caption#1[#2]#3{\par\addcontentsline{\csname
  ext@#1\endcsname}{#1}{\protect\numberline{\csname
  the#1\endcsname}{\ignorespaces #2}}\begingroup
    \small
    \@parboxrestore
    \@makecaption{\csname fnum@#1\endcsname}{\ignorespaces #3}\par
  \endgroup}
\catcode`@=12

\input epsf


\section{Introduction}
\setcounter{footnote}{0}

The possibility of large extra spacetime dimensions has
recently received considerable attention.
There are two fundamentally different ways in which such extra
dimensions might arise.
First, they may appear as extra dimensions felt
by all of the particles and forces of nature, both
gauge and gravitational.  Such extra dimensions are therefore universal,
and apply to all observable physics.
These sorts of large extra dimensions can have important
consequences.  For example, in
Ref.~\cite{DDG},
it was shown that large extra spacetime dimensions of this type
could be used to lower the grand unification (GUT) scale.
This demonstrates that extra universal dimensions have the power
to alter one of the fundamental high energy scales of physics.
Large extra dimensions of this type
also provide a natural way of explaining the fermion
mass hierarchy by permitting the fermion masses to evolve
with a power-law dependence on the energy scale~\cite{DDG}.
Moreover, as first investigated in Ref.~\cite{antoniadis},
large extra universal dimensions can also be used
to induce supersymmetry-breaking via the Scherk-Schwarz
mechanism~\cite{SS}.
Such supersymmetry-breaking scenarios have a number of interesting
signatures~\cite{antoniadis,missusy,AQ,ADS}.
Other phenomenological properties of string theories with large extra
universal dimensions have been discussed in Refs.~\cite{Banks,Ross}.

However, these are not the only types of extra dimensions that might
arise.   For example, there may also be extra dimensions that are
felt only by the gravitational force, with the observable world of
the Standard Model
restricted to a ``brane''.
Such extra dimensions can also play a key role.
For example,
they emerge naturally in describing the
strong-coupling behavior of certain string theories~\cite{HW}.
More recently, it has even been proposed that large extra dimensions of this
type may be used to lower the fundamental Planck scale to the TeV range
and thereby avoid the gauge hierarchy problem~\cite{Dim}.
Such extra dimensions may also be used in a field-theoretic and
string-theoretic
context to transmit supersymmetry-breaking between
four-dimensional boundaries~\cite{Horava,Nilles,Peskin}, and
indeed this leads to a new ``world-as-brane'' perspective which
has been investigated in Refs.~\cite{preSundrum,Dim,Sundrum}.

Finally, large extra dimensions of both types also play an important
role in lowering the fundamental string scale, as first pointed out in
Ref.~\cite{Witten}.  This idea of lowering the string scale was later
dramatically extended to the TeV range in Ref.~\cite{lykken}, and
subsequently pursued in the context of string theories with extra
large dimensions in Refs.~\cite{Dim,henry,DDG,bachas}.

Extra dimensions of both types can be expected to have a profound effect on the
dynamics of the early universe. There are many possible effects which come
into play in the context of a higher-dimensional cosmology~\cite{rocky}. In
addition to the issue of inflation occurring in $D>4$ dimensions, one might
think of the effects of extra dimensions on cosmological phase
transitions,  cosmological  density
perturbations, and  topological defects.

In this paper, we shall consider several aspects of large extra dimensions
as they relate to the dynamics of the early universe.
In doing so, we shall follow two complementary approaches.

First, we shall consider the effects of large extra dimensions through
a field-theoretic analysis.
Most of the applications of field theories are based upon the theory of
phase transitions~\cite{linde}. In particular, the concepts of spontaneous
symmetry breaking in gauge theories~\cite{lee}
and symmetry restoration at high temperatures~\cite{dj} play a
fundamental role. At temperatures above a certain critical temperature, gauge
and/or global symmetries are restored and the order parameter --- usually
the vacuum expectation value of a scalar field --- vanishes. Of particular
interest for cosmology is the nature of the phase transition, whether it
is first-order or not. In most models, this depends upon the mass of the
scalar field. If the phase transition is strongly first-order, the
universe may be dominated by the vacuum energy and undergo a period of
inflation~\cite{revinf}. A first-order phase transition proceeds by nucleation
of
bubbles of the true vacuum, and this dynamics
might provide the local out-of-equilibrium conditions that are a
necessary ingredient for the formation of the baryon asymmetry in the
early universe~\cite{revbau}.
On the other hand, if the phase transition is higher-order or
very weakly first-order, thermal fluctuations may drive the transition.
Spontaneous symmetry-breaking phase transitions may also lead to the
formation of topological defects, which may take the form of domain
walls, cosmic strings, and magnetic monopoles in Grand Unified Theories
(GUTs). These cosmological objects may be either very insalubrious or have
great potential for cosmological relevance.  The latter case is particularly
true for cosmic strings.

In Sect.~2, we will take a field-theoretic approach in order to
explore the role of extra dimensions in
cosmological phase transitions. Specifically, we will study the issue of
restoration of spontaneously broken symmetries above a critical
temperature in the case in which the temperature of the system is larger
than the inverse of the compactification radii. As we shall see,
an interesting distinction
emerges between the critical temperature above which symmetry restoration
takes place in four dimensions, and the corresponding critical temperature
in $D>4$:   the former may be much larger, \ie,
cosmological phase transitions in extra dimensions
can be delayed.  We will also argue that the cosmological phase transitions,
if they happen to take place at temperatures larger than the inverse
radii, cannot be of the first order --- \ie, they do not proceed by
nucleation of critical bubbles.
We will also provide general formulae for the effective potential of the order
parameter at  finite temperature, discuss the applicability of our
approximations, and argue about some possible cosmological implications of
our findings.

Ultimately, however, a field-theoretic analysis of the role of extra
large spacetime dimensions is limited by the fact that higher-dimensional
gauge theories are non-renormalizable, and require the introduction
of ultraviolet cutoffs which in turn signal the appearance of new
physics.
Since string theory is the only known consistent
higher-dimensional theory
which lacks the divergences ordinarily associated with non-renormalizable
field theories, it is natural to consider the corresponding
effects of large extra dimensions in string theory.

There also exists another reason why it is important to consider
the extension to string theory.  As we indicated above,
one of the primary motivations for considering large extra dimensions
is that they can lower the fundamental GUT
and Planck scales~\cite{DDG,Dim}.
However, as discussed in Refs.~\cite{lykken,Dim,henry,DDG,bachas},
such scenarios must ultimately be embedded into reduced-scale
string theories in order to be consistent.
For example, if the GUT and Planck scales are reduced to the TeV range,
then this ultimately requires a TeV-scale string theory as well.
Therefore, the ``stringy'' behavior ordinarily associated with
string thermodynamics will now become important at far lower energies
than previously thought relevant in the discussion of the early
universe, and hence will have heightened significance.

In Sect.~3, therefore, we shall consider some aspects of string
cosmology in the presence of large extra spacetime dimensions.
One crucial issue that arises in string theory and string cosmology
is the role of the Hagedorn transition.  As we will review in Sect.~3.1,
the Hagedorn transition is a phenomenon that arises in any theory
containing an exponentially growing number of states as a function of mass,
and string theory is no exception.  Normally, the Hagedorn
transition does not play a crucial role in string cosmology
because it occurs only at temperatures which are roughly equal
to the string scale, and this is usually taken to be near the Planck scale.
However, if the string scale is now significantly lowered
(perhaps even to the TeV range), then the nature and properties of
the Hagedorn transition become of paramount importance.
Moreover, as we shall see, the presence of large extra dimensions
within a Hagedorn-type framework has a number of interesting cosmological
effects.

Another crucial issue that must be addressed in any discussion of
large extra dimensions is the {\it radius}\/ of these dimensions.
Normally, in string theory, spacetime supersymmetry ensures that
the radius is a modulus --- \ie, that it has a flat potential.
However, as is well-known, thermal effects necessarily break
supersymmetry, and therefore it is possible that thermal effects
can themselves create a potential for the radius which might explain
how compactified radii can become large.  In Sect.~3.2, therefore,
we shall calculate such thermal effects within the framework of Type~I
(open) string theory, and show that finite-temperature effects
might indeed be able to generate and stabilize the desired large radii.

\section{Cosmological Phase Transitions in Field Theory}

We begin by discussing some of the effects of large extra dimensions
using a field-theoretic approach.

\subsection{The general setup}

In any discussion of extra spacetime dimensions, we know
that these extra dimensions must be compactified in order to be
consistent with the observed low-energy world consisting of only four flat
dimensions.  For the sake of simplicity, we shall assume that there is
only one extra
dimension, which is compactified on a circle with fixed
radius $R$ where $R^{-1}$ exceeds presently observable energy scales.
The generalization to more than one extra dimension is straightforward.

The appearance of an extra dimension of radius $R$
implies that a given complex quantum
field $\Phi$ now depends not only on the usual four-dimensional
spacetime coordinates $x$, but also the additional coordinate $y$.
Demanding the periodicity of $\Phi$ under
\begin{equation}
              y~\rightarrow~ y+ 2\pi R
\end{equation}
implies that $\Phi(x,y)$ takes the form
\begin{equation}
           \Phi(x,y)~=~\sum_{n=-\infty}^{\infty}\,\Phi^{(n)}(x)
           \,{\rm exp}\left( iny/R\right)~,
\end{equation}
where $n\in \IZ$. The ``four-dimensional'' fields
$\Phi^{(n)}(x)$ are the so-called Kaluza-Klein modes, and $n$ is the
corresponding Kaluza-Klein excitation number. In general, the mass of each
Kaluza-Klein mode is given by
\begin{equation}
           m_n^2~\equiv~ m_0^2+ \frac{n^2}{R^2}~,
\end{equation}
where $m_0$ is the mass of the zero mode. At energies far below $R^{-1}$,
one expects the extra dimension to be unobservable. However, at energies
or temperatures much larger than $R^{-1}$, excitations of many
Kaluza-Klein modes become possible and the contributions of these
Kaluza-Klein modes must be included in all physical computations. It is
clear that only the lowest-lying Kaluza-Klein modes play an important
role, because the contributions of the very heavy modes are suppressed by
their large masses. In particular, at temperatures $T\gg R^{-1}$, one
expects the relevant number of Kaluza-Klein modes to be $\sim RT$.  This
expectation is confirmed by the explicit computation of physical
quantities such as the critical temperature.

It is important to note that
not every state can have Kaluza-Klein excitations. This complication
arises because it is necessary for the Kaluza-Klein excitations to fall
into representations that permit suitable Kaluza-Klein mass terms to be
formed. This issue is particularly important for chiral fermionic states
which cannot be given a Kaluza-Klein mass. One therefore has two choices at
this
stage:  either the chiral fermionic states do not have Kaluza-Klein
excitations, or chiral-conjugate mirror fermions need to be introduced
to form a massive Kaluza-Klein tower.
We also note that if
the extra dimension is compactified on $S^1/\IZ_2$ (a circle subjected
to the further identification $y\rightarrow -y$),
the Kaluza-Klein excitations can be decomposed into even fields
$\Phi_{+}(x,y)=\sum_{n=0}^\infty\Phi^{(n)}(x)\cos(ny/R)$ and odd fields
$\Phi_{-}(x)=\sum_{n=1}^\infty\Phi^{(n)}(x)\sin(ny/R)$.
Since the appropriate transformation of the fields
under the discrete parity $\IZ_2$ is determined by the interactions,
half of the original Kaluza-Klein theory may be projected out
according to the $\IZ_2$ parity of the fields.
If only the odd tower is left, the zero mode is missing.

\subsection{Computing the one-loop effective potential}

Given this setup, we are now in a position to compute the one-loop  effective
potential $V^{{\rm 1-loop}}(\v)$  for a generic order parameter
$\v$. Let us suppose that our theory contains a set of scalar fields
$\chi_i$ ($i=1,...,n$) which, because of their
interactions with the quantum field $\hat{\varphi}$, accrue a
mass squared in the
background of the classical field $\v=\langle \hat{\varphi}\rangle$
given by
\begin{equation}
           M_i^2(\v)~=~m_i^2+m_i^2(\v)~,
\ee
where $m_i^2$ is a bare mass which does not depend upon the background field.
In four dimensions, the one-loop effective potential at finite temperature
assumes the familiar form~\cite{dj}
\begin{eqnarray}
         V^{{\rm 1-loop}}_{{\rm bos}}(\v)&=&
       \frac{T}{2}\sum_i\: n_i\:\sum_{n=-\infty}^{\infty}\:
             \int\:\frac{d^3 k}{(2\pi)^3}\:{\rm ln}
                 \left[\omega_n^2 +k^2 +M_i^2(\v)\right]\nonumber\\
         &=& T \,\sum_i\:n_i\:\int\:\frac{d^3 k}{(2\pi)^3}\:
          {\rm ln}\left[1-e^{-\beta\sqrt{k^2 +M_i^2(\v)}}\right]~.
\label{bosgen}
\end{eqnarray}
Here $\beta\equiv 1/T$, $n_i$ is the number of degrees of freedom of the field
$\chi_i$,
and the sum is over the Matsubara frequencies $\omega_n=2\pi n  T$.
In passing to the second line, we have explicitly performed the Matsubara sum
and dropped the zero-temperature one-loop term
$\int d^3 k/(2\pi)^3 \sqrt{k^2+M_i^2(\v)}/2$.

Let us now suppose
that there is an extra dimension which contributes a Kaluza-Klein tower
with an extra mass term $\ell^2/R^2$, where $\ell\in \IZ$.
The one-loop effective potential, as seen from the four-dimensional world,
then reads
\begin{equation}
\label{pot}
       V^{{\rm 1-loop}}_{{\rm bos}}(\v)~=~\frac{T}{2}\sum_i\:
       n_i\:\sum_{\ell=-\infty}^{\infty}\:
       \sum_{n=-\infty}^{\infty}\:\int\:\frac{d^3
       k}{(2\pi)^3}\:{\rm ln}\left[\omega_n^2 +k^2
+M_i^2(\v)+\frac{\ell^2}{R^2}
          \right]~.
\end{equation}
Our next step is to rewrite this expression in terms of a Schwinger
proper-time parameter $s$ using the identity
\begin{equation}
   {d\:{\rm ln}\:A\over dA}~=~\int_0^\infty\:ds\:e^{-sA}~.
\end{equation}
After integration over the three-momentum, this yields
\begin{equation}
\label{pot1}
    V^{{\rm 1-loop}}_{{\rm bos}}(\v)~=~-
    \frac{T}{16\pi^{3/2}}\:\sum_i\:n_i\:\int_0^\infty\:{ds\over s^{5/2}}\:
        e^{-s M_i^2(\v)}
      \:\vartheta_3\left(4\pi i T^2 s\right)\:\vartheta_3\left(
    \frac{is}{\pi R^2}\right)
\end{equation}
where the $\Theta$-functions are defined as
\beq
    \Theta^\alpha_\beta (\tau)
       ~=~ \sum_{n=-\infty}^\infty \, e^{2\pi i n\beta} \,e^{\pi i \tau
(n+\alpha)^2}~
\eeq
with
\beq
       \vartheta_1 \equiv \Theta^{1/2}_{1/2}~,~~~~
       \vartheta_2 \equiv \Theta^{1/2}_{0}~,~~~~
       \vartheta_3 \equiv \Theta^{0}_{0}~,~~~~
       \vartheta_4 \equiv \Theta^{0}_{1/2}~.
\label{thetadefs}
\eeq
These $\Theta$-functions have the remarkable property
that
\begin{equation}
       \Theta^\alpha_\beta(-1/\tau) ~=~ \sqrt{-i\tau} ~ e^{-2\pi i
\alpha\beta}~
         \Theta^{-\beta}_\alpha(\tau)
\label{Stransform}
\end{equation}
where one chooses the branch of the square root with non-negative real part.

Let us focus on the limit $RT\gg 1$.
By making the change of variable $s^\prime=4\pi T^2 s$ and subtracting
the $T=0$ part of the one-loop potential, the expression
(\ref{pot1}) becomes
\begin{equation}
\label{pot2}
    \Delta V_{\rm bos} ~=~
        -\frac{T^4}{2}\:\sum_i\:n_i\:\int_0^\infty \,{ds\over s^{5/2}}\,
       e^{-s M_i^2(\v)/4\pi T^2}
           \:\left(\vartheta_3\left(is\right)
      -{1\over\sqrt{s}}\right)\:\vartheta_3\left(
      \frac{is}{4\pi^2 R^2 T^2}\right)
\end{equation}
where we have introduced the notation
\beq
    \Delta V ~\equiv~
       V^{{\rm 1-loop}}(\v)~-
          ~V^{{\rm 1-loop}}(\v)\biggl|_{T=0}~.
\eeq
Here
\beqn
\label{zeropotbos}
     V_{\rm bos}^{\rm 1-loop}(\v)\biggl|_{T=0} &=&
     - {1\over 32\pi^2} \, \sum_i \,n_i\, \int_{\Lambda^{-2}}^\infty
          \, {ds\over s^3} \, e^{-s M_i^2(\varphi_c)}\, \vartheta_3\left(
          {is \over \pi R^2}\right) \nonumber\\
        &=& \sum_i n_i (R\Lambda)\left\{
     {-\Lambda^4\over 80\pi^{3/2}}+{M_i^2 \Lambda^2\over 48\pi^{3/2}}
     -{M_i^4\over 32\pi^{3/2}}+ {M_i^5\over 60\pi\Lambda}
     +{\cal O}(M_i^6) \right\}\nonumber\\
          &&~~~~~~~~+V_{\rm bos}^{D=4}
\eeqn
where $\Lambda$ is an ultraviolet cutoff and
where $V_{\rm bos}^{D=4}$ represents the usual bosonic
contribution to the four-dimensional one-loop Coleman-Weinberg
effective potential~\cite{cw}.
Note that the zero-temperature potential
scales as $R\Lambda$, which is the effective number of Kaluza-Klein
states below the cutoff $\Lambda$.
If we now use the fact that
\begin{equation}
       \vartheta_3(4\pi i\tau) ~\approx~ {1\over \sqrt{4\pi\tau}}
        \, \left[1+{\cal O}\left( e^{-1/4\tau}\right)\right] ~~~~~
          {\rm as}~~ \tau\to 0~
\end{equation}
and perform a high-temperature expansion, $M_i(\v)/T\ll 1$, we find that
the expression (\ref{pot2}) reduces to
\begin{eqnarray}
\label{potbos}
       \Delta V_{\rm bos} &=&\:\sum_i\:n_i\:(RT)\left\{-\frac{3}{2\pi}
       \zeta(5)T^4+\frac{\zeta(3)}{
       4\pi}T^2 M_i^2(\v)\right.\nonumber\\
       &&~~+~ \left. \frac{1}{64\pi}M_i^4(\v)\left[-3 +4\:{\rm ln}(M_i(\v)/T)
       \right]-{M_i^5(\v)  \over 60\pi T}
       \right\}+~{\cal O}(M_i^6) \nonumber\\
       &&~~+~V_{\rm bos}^{D=4}(T)~
\end{eqnarray}
where $\zeta(p)\equiv \sum_{n=1}^\infty~1/n^p$ is the Riemann zeta-function
and where $V_{\rm bos}^{D=4}(T)$ is the usual four-dimensional
finite-temperature effective potential~\cite{dj}.
The first term in this expansion accounts for the total pressure of the
relativistic bosonic particles in the gas.  The fact that the terms of the
expansion are multiplied by the factor $RT$ does not come as a surprise.
At temperatures $T\gg R^{-1}$, there are approximately $RT$ Kaluza-Klein states
which may be treated as massless and which are, therefore, excited in the
thermal bath. This set of states contributes to the effective potential.
By contrast,
the Kaluza-Klein states whose masses exceed the temperature
are essentially decoupled from the thermal bath.

It is also important to note that in the
$RT\gg 1$ expansion of the effective potential, we do not recover any odd
powers of the mass $M_i(\v)$. This is very different from what happens
in $D=4$ field theory at finite temperature, where the infrared limit
$|k|\rightarrow 0$ becomes problematic around $M_i(\v)=0$. At any order of
perturbation theory, the infrared divergence comes from the Feynman
diagrams where the momenta of the particles in the loop correspond to the
$n=0$ Matsubara mode, and give rise to odd powers of $M_i(\v)$.
These odd powers of $M_i(\v)$  play a fundamental role in
the dynamics of cosmological phase
transitions in $D=4$ because their presence induces an energy barrier
which separates the extremum of the scalar potential associated with the
symmetric phase from a local minimum of the broken phase.  At the critical
temperature $T_c$, both phases are equally favored energetically, and at
later times the broken-phase minimum becomes the global minimum. The phase
transition proceeds by nucleation of bubbles of the true vacuum,
signalling a first-order phase transition.

Our findings indicate that for one extra compactified dimension, the
$n=0$ Matsubara frequency mode induces a term ${\cal O}(T M^4_i(\v))$,
which has even powers of $M$. At finite $R$, one also obtains the term
${\cal O}(R M^5_i(\v))$, but this is cancelled by the contribution
from the $T=0$ one-loop effective potential.  This can  be
seen explicitly in (\ref{zeropotbos}) and (\ref{potbos}).
This observation can also be seen
and generalized through a simple scaling argument. In the flat-space limit
$R\rightarrow \infty$, it is sufficient to consider the effective
potential in $D$ spacetime dimensions
\be
\label{deffpot}
       V~=~{T\over 2}\:\sum_n\:\int\: {d^{D-1}k\over (2\pi)^{D-1}}\:{\rm ln}
         (4\pi^2 T^2 n^2 + k^2+M^2)~.
\ee
Formally taking derivatives of the effective potential leads to the
convergent expressions
\be
   {\partial^{N+1}V\over \partial(M^2)^{N+1}}=\cases{
        {\displaystyle {(-1)^N\over 2^{2N+1} \pi^N}T\left[{1\over M^2}
        +2\sum_{n=1}^{\infty}
        {1\over(M^2+4\pi^2T^2 n^2)}\right]} & for $D=2N+1$ \cr
        {\displaystyle {(-1)^N\over 2^{2N+1}}\sqrt{\pi}T\left[{1\over M^3}
       +2\sum_{n=1}^{\infty}
        {1\over(M^2+4\pi^2T^2 n^2)^{3/2}}\right]} & for $D=2N$~. \cr
    }
\ee
Integrating the above expressions for the derivatives and expanding
in powers of $(M/T)$, we find that
\be
  V(M^2)~\sim~\cases{
       {\displaystyle TM^{2N} + TM^{2N} \left({M^2\over T^2}+ {M^4\over T^4}
         + {M^6\over T^6}+...\right)} & for $D=2N+1$ \cr
        {\displaystyle TM^{2N-1} + M^{2N}\left( {M^2\over T^2}+ {M^4\over T^4}
         + {M^6\over T^6}+...\right)} & for $D=2N$~ \cr
        }
\ee
where we have neglected all numerical coefficients in the expansion.
Note that for $D=4$, we obtain the usual cubic term in the four-dimensional
finite-temperature effective potential. However, when the number of extra
dimensions is odd ($D=5,7,9,...$), we see that no odd
powers of $M$ appear in the
high-temperature expansion of the potential.  This agrees with our
explicit calculation in $D=5$ (note that (\ref{deffpot}) implicitly includes
the $T=0$ one-loop effective potential). On the other hand, for even
dimensions ($D=6,8,10,...$), an odd power of $M$ exists in the potential
and specifically arises from the Matsubara zero-mode.  When these even
dimensions are compactified with a finite radius $R$, this term becomes
$(RT)^{D-4} T^4 (M/T)^{D-1}$ where we have inserted a factor of $R^{D-4}$
resulting from the compactification.
Note that the factors of $(RT)^{D-4}$ take into account the effective numbers
of Kaluza-Klein states at temperature $T$.
This term is suppressed relative to the term
$(RT)^{D-4} M^4$ by a factor $(M/T)^{D-5}$ for even dimensions $D>4$.
At finite $R$, there is also the usual $T M^3$ term which
cannot be seen in the $R\rightarrow \infty$ limit. In the limit
$MR\gg 1$, the term $T M^3$ is suppressed by a factor $1/[(RT)^{D-5}(MR)]$
compared to the term
$(RT)^{D-4} M^4$, and for even dimensions $D>4$ this term is suppressed by
$1/(MR)^{D-4}$ relative to the term $(RT)^{D-4} T^4 (M/T)^{D-1}$.
Thus, once again there is no barrier, and we conclude that
for $M\gg R^{-1}$ no barrier is
present in the effective potential in $D>4$ dimensions at high temperature.
This fact  might have dramatic
consequences for the dynamics of phase transitions,
such as the electroweak phase transition,
in the early universe.  We leave this subject for future
investigation.

In an analogous way, we can also compute the contribution to
the one-loop effective potential
from a set of fermion fields $\psi_i$ ($i=1,...,n$) which,
because of their interactions with the quantum field $\hat{\varphi}$, accrue
a mass squared $M_i^2(\v)$ in the background.
For fermions, (\ref{bosgen}) is replaced by
\begin{eqnarray}
         V^{{\rm 1-loop}}_{{\rm fer}}(\v)&=&
       -\frac{T}{2}\sum_i\: n_i\:\sum_{n=-\infty}^{\infty}\:
             \int\:\frac{d^3 k}{(2\pi)^3}\:{\rm ln}
                 \left[\omega_n^2 +k^2 +M_i^2(\v)\right]\nonumber\\
         &=& -T \,\sum_i\:n_i\:\int\:\frac{d^3 k}{(2\pi)^3}\:
          {\rm ln}\left[1+e^{-\beta\sqrt{k^2 +M_i^2(\v)}}\right]~
\label{fermgen}
\end{eqnarray}
where the sum is now over the
shifted (fermionic) Matsubara frequencies $\omega_n=(2 n+1)\pi  T$.
Incorporating the effects of an extra dimension is handled as before.
In the general case of chiral fermions, chiral-conjugate mirror fermions
need to be introduced if the fermions are to have a Kaluza-Klein tower.
However, for simplicity, we shall consider Kaluza-Klein states with
Dirac masses $\ell^2/R^2$.
Let us again consider the limit $RT\gg 1$.
Proceeding just as we did below (\ref{bosgen}),
we obtain
\begin{equation}
\label{potfer}
   \Delta V_{\rm fer} ~=~
   +\frac{T^4}{2}\:\sum_i\:n_i\:\int_0^\infty\,
       {ds\over s^{5/2}} ~
     e^{-s M_i^2(\v)/4\pi T^2} \,
   \left(\vartheta_2\left(is\right)-{1\over\sqrt{s}}\right)\:\vartheta_3\left(
   \frac{is}{4\pi^2 R^2 T^2}\right)~.
\end{equation}
This is the fermionic analogue of (\ref{pot2}).
In the limit $RT\gg 1$, we can
use (\ref{Stransform}) to evaluate this integral, obtaining
\begin{eqnarray}
\label{potferoneloop}
    \Delta V_{\rm fer}
      &=&\:\sum_i\:n_i\:(RT)\:\left\{-\frac{45}{32\pi}\zeta(5)T^4+
      \frac{3\zeta(3)}{16\pi}T^2
      M_i^2(\v)\right. \nonumber\\
      &&~~~~~~-~\left. \frac{\ln 2}{16\pi}M_i^4(\v)+{M_i^5(\v)\over 60\pi T}
      \right\}+{\cal O}(M_i^6)~~+~ V_{\rm fer}^{D=4}(T)~
\end{eqnarray}
where $V_{\rm fer}^{\rm 1-loop}(\v)|_{T=0}=
-V_{\rm bos}^{\rm 1-loop}(\v)|_{T=0}$ and
where $V_{\rm fer}^{D=4}(T)$ is the fermionic contribution
to the four-dimensional finite-temperature effective potential~\cite{dj}.
Note, in particular, that the fermionic contribution to the squared mass
term carries the same sign as the bosonic piece in (\ref{potbos}). This
is a typical feature of high-temperature field theories where, even starting
with a supersymmetric theory, supersymmetry is broken by finite-temperature
effects.  Also note that although the sign of the quartic term $\v^4$ in
(\ref{potferoneloop}) is
negative, the potential is not destabilized because the tree-level quartic
term continues to dominate in the limit $\lambda (RT)\simlt 1$  where
$\lambda$ is the coefficient of the quartic term in $V(\varphi_c)$.
This happens to be the limit in which the one-loop computation is
reliable, as we shall discuss in the next subsection.

Finally, for completeness, let us discuss some issues in the context
of Scherk-Schwarz supersymmetry breaking by compactification.  This has some
features that are similar to those of the
finite-temperature calculation.
At $T=0$, combining the contributions from bosons and fermions leads
to the the one-loop effective potential
\begin{eqnarray}
\label{sspot}
  V_{\rm SS}^{\rm 1-loop}(\v)&=&{1\over 2}\sum_i\: n_i\:
       \sum_{n=-\infty}^{\infty}\:\int\:\frac{d^4 k}{(2\pi)^4}\:
       \Biggl\{{\rm ln}\left[k^2 +M_i^2(\v)+{n^2\over R^2}\right]\nonumber\\
        && ~~~~~~~~~~-{\rm ln}\left[k^2 +M_i^2(\v)+{(n+1/2)^2\over
R^2}\right]\Biggr\}
                         \nonumber\\
    &=& - {1\over 32\pi^2}\sum_i n_i\int_0^\infty {ds\over s^3}\, e^{-s M_i^2}
       \,\left[\vartheta_3({i s\over\pi R^2})
             -\vartheta_2({i s\over\pi R^2})\right]~.
\end{eqnarray}
This result is free of ultraviolet divergences, as expected.
In the limit $RM\ll 1$, we can perform the change of variable
$s^\prime=(s/\pi R^2)$ to find that the contribution from the
Kaluza-Klein states to the one-loop effective potential reads
\begin{eqnarray}
\label{zerot}
     V^{\rm 1-loop}_{\rm SS}(\v)&=&
     \sum_i\:n_i\:\left\{-\frac{93}{1024\pi^6}\frac{\zeta(5)}{R^4}
     +\frac{7\zeta(3)}{128\pi^4}\frac{M_i^2(\v)}{R^2}\right.\nonumber\\
       &&~~~~+~ \left.\frac{M_i^4(\v)}{128\pi^2}
      \left[-3 +4\:{\rm ln}(\pi M_i(\v)R)\right]\right\}
      +{\cal O}(M_i^6)~.
\end{eqnarray}
The first term in (\ref{zerot}) is a Casimir force term, which
is expected to arise in such a non-supersymmetric model.
More interestingly, however, we observe from (\ref{zerot})
that the squared mass term receives a finite contribution
which scales as $1/R^2$. If the order parameter $\v$
is associated with the low-energy Standard Model Higgs field,
this implies (by naturalness arguments) that $1/R$ has to
be smaller than about 10-100~TeV.
This raises the interesting possibility of
breaking the Standard-Model gauge group via the contributions of the
Kaluza-Klein states if the $1/R^2$ term happens to carry a negative sign.
This is possible in more general Scherk-Schwarz compactification
scenarios.\footnote{
      In this connection, note that
     in compactifying from five dimensions to four dimensions,
     there are two symmetries which can be exploited for the Scherk-Schwarz
breaking.
     The first is the fermion number operator $(-1)^F$, which
           gives $\sum_i n_i = n_V+n_H$  where $n_V$  ($n_H$) is the number of
             five-dimensional $N=2$ vector (hyper-) multiplets.
      The second symmetry, by contrast,
           is the $\IZ_2$ $R$-parity, which gives
           $\sum_i n_i= n_V-n_H$.  It is the second symmetry which can
           lead to a negative contribution for the $1/R^2$ term.  Similar
       considerations can also be found in
      Ref.~\cite{marianotalk}.}

At finite temperature, we can perform the explicit sum over the Matsubara
frequency modes that appear in (\ref{sspot}) after compactifying the
time coordinate.  Performing a high-temperature expansion in the limit
$RT\gg 1$ then leads to the result
\begin{eqnarray}
     V^{\rm 1-loop}_{\rm SS}(T)&=&
     \sum_i\:n_i\:(RT)\left\{-\frac{93\zeta(5)}{32\pi} T^4
     +\frac{7\zeta(3)}{16\pi}T^2 M_i^2(\v)\right.\nonumber\\
       &&~~~~+~ \left.\frac{M_i^4(\v)}{64\pi}
      \left[-3 +4\:{\rm ln}(M_i(\v)/2T)\right]\right\}
      +{\cal O}(M_i^6)~.
\end{eqnarray}
Again we notice the overall factor $RT$  which is the effective number
of Kaluza-Klein states. In addition, we find that to leading order,
the result is simply
the sum of the terms (\ref{potbos}) and (\ref{potferoneloop}).  This
is expected,  for in the high-temperature expansion, the
supersymmetry-breaking mass difference between the bosons
and fermions is negligible.

\subsection{Multi-loop corrections}

Let us now consider what happens when
multi-loop corrections are included in the computation
of the effective potential. For this purpose, we shall assume that our
theory is a simple
$(\lambda/4)\hat{\varphi}^4$-theory. In such a case, the corrections to the
tree-level potential are provided by the quantum and/or thermal
excitations of the $\varphi$-field itself and, in particular, by its
Kaluza-Klein tower. As opposed to large-angle scattering processes,
forward-scattering processes do not alter the distribution function of
particles traversing a gas of quanta;  they instead simply modify the
dispersion
relation. Forward scattering is manifest, for example, as ensemble and
scalar-background corrections to the masses of the particles in the
plasma. In other words, when they propagate, the particles in equilibrium
in the thermal gas acquire a plasma mass $\delta m(T)$ through forward
scatterings. Now, if the thermal environment is at some temperature $T$,
only those excitations which are lighter than approximately $T$ may be in
thermal
equilibrium and present in the plasma. Other thermal excitations with masses
much larger than $T$ are decoupled from the thermal bath, and do not
alter the potential.

Given the expression (\ref{potbos}),
the one-loop plasma mass of the $\varphi$-quanta
is easily found to be $\delta m^2(T)\sim
\lambda (RT) T^2$. This means that for $\lambda (RT)\gg 1$,
the quanta responsible for the one-loop correction to the potential of the
order parameter $\v$ are in fact much heavier than $T$.  This means that
they should decouple, and give no contribution to the effective
potential!  In other
words, the one-loop high-temperature expansion that we used to derive
(\ref{potbos}) breaks down in a certain range of the parameters, \ie,
$\lambda (RT)\gg 1$, which is where the one-loop effective potential
should receive large contributions from two- and higher-loop orders
of perturbation theory.  These contributions will be
even larger than the one-loop contribution.
Perturbation theory is therefore invalid unless a proper resummation is done.

Given this observation, we see that
in order to obtain more accurate
information about the issue of cosmological phase transitions in $D>4$
dimensions, we have to study an infinite series of diagrams in perturbation
theory. This is exactly analogous to what happens in a simple
$\lambda\hat{\varphi}^4$ theory in equilibrium at finite temperature in $D=4$,
where the leading contributions to the effective potential in the
infrared region come from the daisy and superdaisy multi-loop
graphs~\cite{dj}.

In order to deal with this problem, we need a self-consistent loop
expansion of the effective potential in terms of the {\it full}\/ propagator.
Such a technique was developed some time ago by Cornwall, Jackiw, and Tomboulis
(CJT) in their effective-action
formalism for composite operators~\cite{cjt}.
In the rest of this section, we will consider a scalar field
theory where $\hat{\varphi}$ transforms as a vector under the action of
$O(N)$, \ie,  $\hat{\varphi}^2 = \hat{\varphi}_a \hat{\varphi}^a$
with $a=1,...,N$, with a potential given by
$V(\varphi)=(\lambda/4)(\hat{\varphi}^2)^2$. For the sake of simplicity, we
will assume that there are no fermions in the theory.

We consider a generalization $\Gamma[\varphi_c,G]$ of the
usual effective action which depends not only on $\varphi_c(x)$, but also
on $G(x,y)$, a possible expectation value of the  time-ordered
product $\langle T\varphi(x)\varphi(y)\rangle$. The
physical solutions must satisfy the stationary conditions
\begin{equation}
\label{gg}
   \frac{\delta\Gamma[\varphi_c,G]}{\delta\varphi_c(x)}~=~ 0~,~~~~~~
   \frac{\delta\Gamma[\varphi_c,G]}{\delta G(x,y)}~=~ 0~.
\end{equation}
The conventional effective action $\Gamma[\varphi_c]$ is given by
$\Gamma[\varphi_c,G]$
at the solution $G_0(\varphi_c)$ of (\ref{gg}).
In this formalism, it is possible to sum a large class of ordinary
perturbation-series diagrams that contribute to the effective action
$\Gamma[\varphi_c]$, and the gap equation which determines the form of the
full propagator is obtained by a variational technique.

We now apply the CJT formalism in the limit of large
$N$ when the next-to-leading terms can be exactly summed.
At each order, we keep only the term dominant in $N$ for large values
of $N$. This allows us to resum the series of the leading
multi-loop diagrams exactly and to solve the gap equation for the full
propagator without any approximation.

In order to obtain a series expansion of the effective action, we
introduce the functional operator
\begin{equation}
     D^{-1}_{ab}(\varphi_c,x,y)
     ~=~\frac{\delta^2\,I}{\delta\varphi_{c,a}(x)\,\delta
     \varphi_{c,b}(y )}
\end{equation}
where $I$ is the classical action. The required series obtained by
CJT is then~\cite{cjt}
\begin{equation}
     \Gamma[\varphi_c,G]~=~
   I(\varphi_c)+\frac{1}{2}\:{\rm Tr}\:{\rm ln} \:D_0\:G^{-1}
     + \frac{1}{2}\:{\rm Tr}\:\left[D^{-1}\:G-1\right]+\Gamma_2[\varphi_c,G]~,
\end{equation}
where $D_0^{-1}=-\left(\partial_\mu\partial^\mu\right)\delta_{ab}\delta^
4(x,y)$.  Here $\Gamma_2[\varphi_c,G]$ is the sum of all
two-particle-irreducible vacuum graphs in the theory, with vertices defined
by the classical action with shifted fields $I[\varphi_c+\varphi]$ and
propagators set equal to $G(x,y)$.

Previous calculations show that among the multi-loop graphs contributing to
the effective potential in the $O(N)$-theory, only the daisy and
superdaisy diagrams survive in the large-$N$ limit~\cite{limit}. This
enables us to consider in $\Gamma_2[\varphi_c,G]$ only the graph of ${\cal
O}(\lambda)$. This is essentially the Hartree-Fock approximation,
which is known to be exact in the many-body version of our large-$N$ limit.

\begin{figure}[ht]
\centerline{ \epsfxsize 4.0 truein \epsfbox {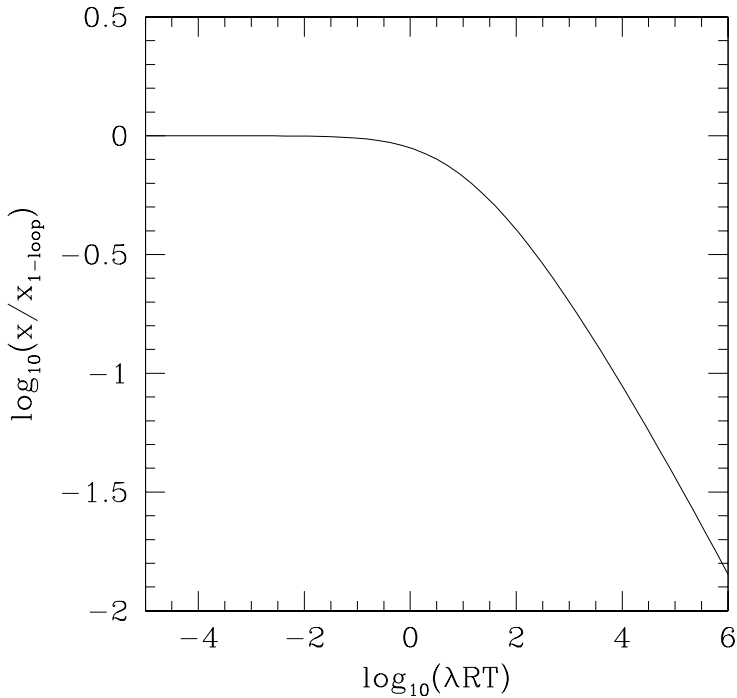}}
\caption{The ratio $x/x_{{\rm 1-loop}}$ as a function of $\lambda RT$.}
\label{figureone}
\end{figure}

It turns out to be more convenient to concentrate on
the effective masses rather  than on the effective potential.
By stationarizing the effective action $\Gamma[\varphi_c,G]$ with respect
to $G_{ab}$, we  obtain the gap equation
\begin{equation}
\label{p}
   G_{ab}^{-1}(x,y)~=~D_{ab}^{-1}(x,y)+3\lambda\left[\delta_{ab
   }\:G_{cc}(x,x)+2\:G_{ab}(x,x)\right]\:\delta^4(x,y)~.
\end{equation}
This equation is exact in the limit of large $N$, and contains all the
information about the dominant-$N$ contributions to the full
propagator. Indeed, the exact Schwinger-Dyson  equation reduces to
(\ref{p}) for large $N$.
Next, we Fourier-transform (\ref{p}) and take $\varphi_c=0$.
The gap equation then reads
\begin{equation}
    M^2~=~\frac{3\lambda}{2\pi}(RT)T^2\left[\frac{M}{T}\:{\rm
    Li}_2\left(e^{-M/T}\right)+{\rm Li}_3\left(e^{-M/T}\right)\right]~
\label{ge}
\end{equation}
where Li$_n(x)\equiv\sum_{k=1}^\infty x^k/k^n$ are the polylogarithm
functions.
As we discussed above, in
the limit $\lambda(RT)\ll 1 $ we expect that the value of $M^2$ solving
the gap equation is well-approximated by the one-loop mass $M_{{\rm
1-loop}}^2=(3\zeta(3)/2\pi)\lambda(RT)T^2$. On the other hand, for
$\lambda(RT)\gg 1 $, we expect the solution $M^2$ to be quite different
from $M_{{\rm 1-loop}}^2$. Our expectations are indeed confirmed in
Fig.~\ref{figureone},
where we have defined $x\equiv M/T$ to be the solution of the gap
equation (\ref{ge}) and plotted the value of $x$ normalized to
$x_{{\rm 1-loop}}=\sqrt{3\zeta(3)\lambda(RT)/2\pi}$.
Note, in particular, that at large values of $\lambda(RT)$, the value of $x$ is
much smaller than the value indicated by the one-loop computation,
indicating that the plasma mass is screened by higher-order corrections.
This means that the usual perturbative expansion fails for
$\lambda(RT)\simgt 1$, and has to be replaced by an improved
perturbative expansion where an infinite number of diagrams are resummed
at each order in the new expansion.

\subsection{Implications of large extra dimensions}

We have seen that in $D>4$ dimensions, the effective potential
of a given order parameter at high temperature has some peculiar features
which are not present
in the case in which only four dimensions are experienced. This means
that the dynamics of  cosmological phase transitions taking place in the
early universe
when the Compton wavelength $\sim T^{-1}$ of the thermal excitations
was still smaller than the length scale of the extra dimension(s)
is different from the usual dynamics in four dimensions.

To simplify our discussion, let us consider again the simplest scalar
field theory with the potential
\begin{equation}
       V(\v)~=~-\frac{\mu^2}{2}\v^2+\frac{\lambda}{4}\v^4
\end{equation}
where $\mu^2$ is a positive bare mass term. The vacuum expectation value of
the scalar field in the present vacuum is therefore
$\langle \v\rangle=\mu/\sqrt{\lambda}$. We know, however, that in
$D=4$ and at very high temperature, the bare mass $\mu$ receives a
temperature-dependent correction $\delta m^2(T)=\lambda T^2/4$.  As a
result, for temperatures higher than the critical temperature
\begin{equation}
\label{d4}
       \left(T_c\right)_{D=4}~=~ 2\frac{\mu}{\sqrt{\lambda}}~,
\end{equation}
the vacuum expectation value of the scalar field vanishes. This is the
signal of a phase transition.

Let us now suppose that there is an extra dimension which opens up at a
certain length scale $R$, and let us follow the dynamics of the system.
Since at energies smaller than $R^{-1}$ one should recover the low-energy
effective theory in four dimensions, it is reasonable to assume that
$\mu$ is smaller than $R^{-1}$.

When the thermal gas was extremely hot, such that $RT\gg 1$, the universe
was effectively five-dimensional.  Now, if there appear $T=0$ corrections
to the effective potential of the form (\ref{zerot}) --- and if
the correction ${\cal O}(1/R^2)$ to the bare squared mass $-\mu^2$
is negative --- then it is clear that the phase transition will occur at a
temperature $T_c={\cal O}(R^{-1})$,  independently  of the
value of $\mu$. This is already an interesting result if we believe, for
instance, that there is an extra dimension at the TeV-scale and the order
parameter is the Standard-Model Higgs field.
Under these circumstances,
the electroweak phase transition will be very  different from what is
usually expected.

If, on the other hand, the corrections of the form (\ref{zerot}) are not
present (\eg, as would arise without Scherk-Schwarz supersymmetry-breaking),
we can easily convince ourselves that for $\lambda RT\simgt
1$, the plasma mass $\delta m(T)$ is much smaller than the value suggested
by the one-loop analysis, but nevertheless too large for the phase
transition to occur.

When the universe cools down to values of the temperature $T\simlt
(\lambda R)^{-1}$, but still larger than $R^{-1}$, the phase transition
may occur when the plasma squared mass becomes larger than the negative
bare squared mass $-\mu^2$. This takes place at the critical temperature
\begin{equation}
\label{d5}
    \left(T_c\right)_{D=5}~=~\left(\frac{2\pi}{3\zeta(3)}\frac{\mu^2}
                           {\lambda R}\right)^{1/3}~.
\end{equation}
This estimate is valid as long as $T_c R\simgt 1$ and the high-temperature
expansion is valid. This translates into the bound
$\sqrt{\lambda}\simlt \mu R\simlt \lambda^{-1}$, which is not very
stringent if $\lambda\simlt 1$. In addition, one should also be aware of the
power-law running of the four-dimensional couplings~\cite{DDG},
since this will
affect the determination of the critical temperature. Indeed the
critical temperature (\ref{d5}) satisfies $RT_c\simlt 1/\lambda$, and for
$\lambda \simlt 1$ the power-law running caused by the large number of
Kaluza-Klein states can drastically affect the couplings
 in the tree-level potential.
Of course, this issue should be addressed in more realistic theories
than we are considering here.

Note that the ratio between the critical temperatures
(\ref{d5}) and (\ref{d4}) is
\begin{equation}
     r~\equiv~ \frac{\left(T_c\right)_{D=5}}{\left(T_c\right)_{D=4}}~\approx~
     0.6\left(\frac{\lambda^{1/2}}{\mu R}\right)^{1/3}~,
\end{equation}
and lies in the range $\sqrt{\lambda}\simlt r\simlt 1$. We thus see
that the effect of the extra dimension ---
besides preventing a phase transition from being first-order ---
is to delay the instant  at which the phase
transition occurs. This is not  surprising,  because a large
fraction of the Kaluza-Klein tower now contributes to the plasma mass
squared, increasing it by a factor $\sim RT$.

\subsection{Other features}

Even though these results have been obtained for a toy model, we expect
these features to be present in more realistic
theories.  They should therefore have a
profound impact on our understanding of the early universe, and  many aspects
of early-universe cosmology
should be now reconsidered  under the supposition that
the universe might experience extra dimensions at early epochs.
Here, we shall briefly outline some of these features.

We know that the monopole problem is one of the
central issues in modern astroparticle physics. The problem of monopoles
is especially serious since it is generic to the idea of GUTs where the
GUT gauge group is broken via the Higgs mechanism.\footnote{
      Note, however, that in higher dimensions~\cite{DDG},
      the GUT symmetry may be broken via an alternative mechanism
       involving orbifolds (\eg, Wilson lines).  In such cases,
       GUT monopoles may have different properties~\cite{WW}.}
The production of magnetic monopoles  in cosmological GUT phase transitions by
the Kibble mechanism seems almost unavoidable, and is very much akin to the
mechanism for the production of various defects in ordinary laboratory
phase transitions. In the more familiar $D=4$ cosmology, approximately one
monopole per horizon should arise at the GUT phase transition, so that the
resulting monopole-to-entropy ratio is expected to be of the order of
$n_M/s\sim(T_c/M_{{\rm Pl}})^3$, where $M_{{\rm Pl}}\approx 1.2\times 10^{19}$
GeV is the Planck mass. Barring significant monopole-antimonopole
annihilation,  entropy production, or the presence of large global
charges at early epochs which may prevent
the phase transition~\cite{charge1,charge2},
the relic monopole density today is unacceptable.
Let us suppose, however, that the energy scale of
the extra dimension is close to the GUT scale,
so that the GUT phase transition is actually occurring in five dimensions.
A na\"\i ve estimate then leads to a monopole-to-entropy ratio
\begin{equation}
       \frac{n_M}{s}~\sim~ \left(\frac{R^{1/2}T_c^{7/2}}{
       M_{\rm Pl}^3}\right)_{D=5}~,
\label{ratio}
\end{equation}
where we have used the fact that the entropy density scales like $R T^4$ and
that one expects the formation of one monopole per horizon volume. This ratio
(\ref{ratio}) may be much smaller than the corresponding one in $D=4$.
Even though the formation of magnetic monopoles in extra-dimensional
cosmology needs to be addressed more rigorously before any firm conclusion
can be drawn, our estimates seem to suggest that the monopole problem may be
ameliorated in the scenario depicted in this paper.

Other issues include the formation of cosmic strings in $D>4$ dimensions
as the result of the spontaneous breaking of abelian symmetries;
the possibility of extreme supercooling and
a subsequent period of inflation if the energy density of the plasma
becomes smaller than the vacuum energy density associated with a scalar
(inflaton) potential;  and the dynamics of the electroweak phase transition if
the extra dimensions open up at the TeV-scale.


\section{Extension to string theory}
\setcounter{footnote}{0}

In the previous section, we studied the behavior of higher-dimensional
phase transitions using a field-theoretic approach.  However, such
an approach ultimately faces an important limitation.  As we mentioned
in the Introduction, the fact that
such higher-dimensional gauge theories are non-renormalizable implies
that their properties
depend on ultraviolet cutoffs which in turn signal the appearance of new
ultraviolet physics.  String theory is the only known consistent
higher-dimensional theory
which lacks the divergences ordinarily associated with non-renormalizable
field theories.  In this section, therefore, we shall
consider an extension of our analysis to string theory.

There also exists another reason why it is important to consider
the extension to string theory.  As mentioned in the Introduction,
one of the primary motivations for considering large extra dimensions
is that they can lower the fundamental
GUT scale~\cite{DDG} and Planck scale~\cite{Dim}.
However, as discussed in Refs.~\cite{lykken,Dim,henry,DDG,bachas},
such scenarios must ultimately be embedded into reduced-scale
string theories in order to be consistent.
For example, if the GUT and Planck scales are reduced to the TeV range,
then this ultimately requires a TeV-scale string theory as well.
Therefore, the ``stringy'' behavior ordinarily associated with
string thermodynamics will now become important at far lower energies
than previously thought relevant in the discussion of the early
universe, and hence will have heightened significance.

In this section, we shall focus on two such ``stringy'' effects.
The first of these concerns the Hagedorn transition, while the
second concerns the possible generation and stabilization of a
large radius of compactification due to thermal effects.

\subsection{The Hagedorn phenomenon:\\
  limiting temperature vs. phase transition}

One of the most profound differences between string theory
and field theory is the presence of an exponentially growing number
of string states as a function of mass.
These states arise as string oscillator modes which are not
present in a theory in which the fundamental degrees of freedom
are point particles.
As first pointed out in the 1960's by Hagedorn~\cite{Hagedorn},
theories with exponentially growing numbers of states
exhibit a remarkable phenomenon, namely a critical temperature
beyond which the thermodynamic partition function (and indeed
all subsequent thermodynamic quantities) cannot be defined.
This critical temperature is called the Hagedorn temperature,
and can be interpreted either as a limiting temperature or as the location
of a phase transition.   It turns out that this ultimately
depends on
the details of the physical system in question, and will be discussed
in detail below.
Thus, the nature of the Hagedorn phenomenon is ultimately one of the focal
points of any discussion of string theory at finite temperature.

Let us begin by briefly reviewing some of the aspects
of the Hagedorn transition in arbitrary numbers of uncompactified dimensions
$D$.  While the situation in the critical dimension $D=10$ is well-understood,
there is apparently some confusion in the literature regarding the effects
caused by the compactification to $D<10$ and the proper treatment of the
corresponding
Kaluza-Klein excitations.  This will be particularly important for theories
with
large radii of compactification.  Therefore, one of our aims in this section
will
be to resolve these discrepancies.

Let us begin, as in the previous section, by recalling the general expression
for
the free energy at finite temperature in $D$ spacetime dimensions
\beq
    \ln Z ~\sim~ \int_0^\infty dM ~\rho(M)~ \int d^{D-1} k \,
      \ln \left( { 1+ e^{-\beta \sqrt{k^2+M^2}}\over
          1- e^{-\beta \sqrt{k^2+M^2}}} \right)~.
\label{freeenergy}
\eeq
This expression is the $D$-dimensional analogue of (\ref{bosgen}) and
(\ref{fermgen}), where we have assumed a supersymmetric configuration
of bosonic and fermionic states whose density
is given by $\rho(M)$ at mass $M$.  As usual, the total energy of any state
with mass $M$ is given by $E^2= k^2+M^2$, and we have neglected (and
will continue to neglect) overall numerical coefficients.
Note that (\ref{freeenergy}) is simply the expression for
the logarithm of the macrocanonical partition function $Z$, as indicated;
this is related to the potentials $V$ discussed in Sect.~2 via $V= -T\ln Z$.
Also note that unlike the situation in field theory, where we considered
the dependence of $M$ on a background field $\varphi_c$, in string theory
are forced to set $\varphi_c=0$ and treat $M$ as a free parameter.

In our analysis in Sect.~2, we considered only a discrete set of
bosonic or fermionic states, so that $\rho(M)$ was essentially a
delta-function $\delta(M-M(\varphi_c))$.
However, in string theory, we have a much more complicated set of
states which consist of not only Kaluza-Klein modes, but also
string oscillator and winding modes.
As a result, $\rho(M)$ takes the more complicated form
\beq
         \rho(M) ~\sim~ a M^{-b} e^{c M} ~~~{\rm as}~ M\to\infty~.
\label{asymdensity}
\eeq
Here $(a,b,c)$ are presumed to be constant, positive coefficients
whose values depend on the particular system under study.
It is this change in $\rho(M)$ which leads to the important difference
between field theory and string theory.
As we shall see, the parameter $c$ ultimately determines the Hagedorn
temperature of the system, while
the parameter $b$ determines whether
this temperature is to be interpreted as a limiting temperature or
as the site of a phase transition.

Note that for the purposes of this analysis, we are not dealing with a
full string theory.
Rather, we are implicitly dealing with a gas of particles whose properties
(such as the density of states) match the individual modes of the string.
This approximation
will be sufficient for our purposes.

Type~I, Type~II, and heterotic strings all have critical dimensions $D=10$,
and the result (\ref{freeenergy}) applies directly in this case.
The density given in (\ref{asymdensity}) then includes the contributions
from only string oscillator states, since there are no Kaluza-Klein
or winding-mode states resulting from compactification.  However,
once we compactify to spacetime dimensions $D<10$, we must
properly incorporate the contributions of Kaluza-Klein and winding-mode states.
There are two equivalent ways in which this can be done.
The first way, as in the previous
section, is to replace the momentum integrations in (\ref{freeenergy})
that correspond to compactified directions
with discrete summations over Kaluza-Klein and winding modes.  This
then leads, as before, to
products of $\vartheta$-functions in the integrand, and we should continue
to demand that $\rho(M)$ include the contributions of string oscillator states
only.  However, in string
theory it turns out to be simpler to choose a second method:  we can
neglect the contributions from Kaluza-Klein states to the momentum integration
altogether, and simply incorporate their effects in a string calculation
of $\rho(M)$.  It turns out that this changes the value of $b$ without
affecting the value of $c$.
These methods are ultimately equivalent because the Kaluza-Klein
$\vartheta$-functions in the momentum integrand
effectively shift the value of $b$ in $\rho(M)$ by a $D$-dependent amount,
and this amount can be most easily calculated using the underlying
conformal symmetry of the string directly.  Therefore, in this section,
we shall {\it neglect the
contributions of the Kaluza-Klein states in the momentum integral,
and compensate for this by including their effects in the value of $b$.}
This is an important point which has been missed in several prior analyses
of the string Hagedorn transition in $D<10$ dimensions.
We will also follow the same procedure for the string winding states, which
have no analogue in a field theory based on point particles.

With this understanding, let us now proceed to evaluate (\ref{freeenergy}),
taking $D$ to represent the number of {\it uncompactified}\/ dimensions only.
 From this result, we will then be able to calculate the internal energy $U(T)$
as well as other thermodynamic quantities.
We shall follow parts of the approaches
outlined in Refs.~\cite{old,bowick,tye,alvarez}.
Because we are interested in the high-temperature behavior,
the contributions from the extremely massive string states
dominate.
Therefore, we can Taylor-expand the logarithm, obtaining
\beq
    \ln Z ~\sim~ \int_0^\infty dM ~\rho(M)~ \int d^{D-1} k ~
        e^{-\beta \sqrt{k^2+M^2}} ~.
\eeq
Next, we perform the momentum integrations, obtaining
\beq
    \ln Z ~\sim~ \int_0^\infty dM ~\rho(M)~ \beta^{1-D/2}\, M^{D/2}\,
         K_{D/2}(\beta M)~.
\eeq
Here $K_{D/2}(z)$ is the modified Bessel function of third kind, with
asymptotic behavior $ K_\nu(z) \sim z^{-1/2} \,e^{-z}$ as $|z|\to \infty$.
We thus obtain
\beqn
    \ln Z &\sim& \int_0^\infty dM ~\rho(M)~ (M/\beta)^{(D-1)/2}\,
         e^{-\beta M}~\nonumber\\
      &\sim&  \int_0^\infty dM
                 ~M^{-b+(D-1)/2}\, \beta^{(1-D)/2}\, e^{-(\beta-c) M}~
\label{semifinal}
\eeqn
where we have substituted the density (\ref{asymdensity}) into the last line.

The divergence of this integral at the $M\to 0$ endpoint is unphysical,
reflecting the failure of the asymptotic form (\ref{asymdensity}) to
properly reflect the true number of physical states in the $M\to 0$ limit.
For a proper treatment, a more precise functional form should be used
in this limit;  methods for deriving such
forms can be found in Ref.~\cite{Hardy}.
What concerns us here, however, is the opposite extreme
as $M\to \infty$.  Here the asymptotic form (\ref{asymdensity}) is presumed
to be accurate, whereupon we see that the partition function $Z(\beta)$
necessarily
diverges unless $\beta > c$.  This then defines the critical (Hagedorn)
temperature, given by
\beq
              T_H ~\equiv~ c^{-1} ~.
\eeq
For $T< T_H$, the partition function is finite and the corresponding
thermodynamic
properties can be defined without difficulty.  At $T=T_H$, however, this
description
based on the canonical ensemble fails,
and one must resort to a more fundamental description of the physics
(\eg, one based on the microcanonical ensemble)
in order to determine the nature of this transition.

However, as indicated above,
one clue can already be determined directly from the
canonical ensemble.
It is possible to study the behavior of the free
energy $F(T)$ and the internal energy $U(T)$
as functions of the temperature as we approach the Hagedorn
transition from below.
If these thermodynamic quantities also diverge, then
an infinite amount of energy would be required to propel the system past the
Hagedorn temperature.  In such cases, the Hagedorn temperature is
a true limiting temperature of the system.
On the other hand, if these thermodynamic quantities remain finite as $T\to
T_H$,
then infinite amounts of energy are not required, and $T_H$ is more
appropriately
interpreted as the site of a phase transition.

In order to derive the conditions that distinguish between these
two cases, let us first consider the free energy itself
and take $T\to T_H$ (or $\beta\to c$) in (\ref{semifinal}).
The exponential term then cancels, and we are left with
\beq
     \ln Z ~\sim~ \int_0^\infty dM  ~M^{-b+(D-1)/2} ~.
\eeq
This has an ultraviolet divergence for $b\leq (D+1)/2$, with
the divergence becoming logarithmic when this inequality is saturated.
Similarly, the internal energy
  $U(T)\equiv - \partial \ln Z/\partial \beta$
and the entropy
$S(T)\equiv \beta^2 \partial (\beta^{-1}\ln Z)/\partial \beta$
have an
ultraviolet divergence for $b\leq (D+3)/2$, and the specific heat
$c_V\equiv \beta^2 \partial^2 \ln Z/\partial \beta^2$ has an ultraviolet
divergence for $b\leq (D+5)/2$.
These results agree with those found in Refs.~\cite{tye,alvarez}.
In each case, we then find that the relevant thermodynamical quantity
$X(T)$ diverges as
\beq
         X(T) ~\sim~ \cases{
          |\ln ( T_H-T)|  & for $b= b_{\rm crit}$ \cr
          ( T_H-T)^{b-b_{\rm crit}}  & for $b< b_{\rm crit}$ \cr}
\eeq
where $b_{\rm crit}= (D+1)/2+n$,
with $n$ denoting the number of $\beta$-derivatives of $\ln Z$ necessary
to produce $X(T)$.
We conclude that if $b\leq (D+3)/2$, it takes an infinite amount of energy
to raise the temperature of the system past $T_H$ --- \ie, in such cases,
$T_H$ is to be interpreted literally as a physical limiting temperature.
By contrast, for $b> (D+3)/2$, only a finite amount of energy is needed,
whereupon $T_H$ is more appropriately interpreted as a site of a phase
transition.

Let us now consider the values of $b$ and $c$ that arise in string theory.
It is here that we shall have to be careful to properly incorporate the
effects of the Kaluza-Klein states (and winding-mode states)
resulting from compactification.

First, strictly speaking, string theory provides
us not with a density of states $\rho(M)$, but rather a discrete
set of energy levels characterized an oscillation number $n$
and a corresponding number of states $g_n$.
In general, these degeneracies $g_n$
take the asymptotic form
\beq
             g_n ~\sim~ A n^{-B} e^{C\sqrt{n}}~~~{\rm as}~ n\to\infty~
\label{stringasymform}
\eeq
where once again $(A,B,C)$ are positive constants which depend on the
particular string theory in question.
In order to relate (\ref{stringasymform}) to (\ref{asymdensity}),
we need to know how the pure number $n$ relates to the spacetime
mass $M$.  In general, this relation depends on the type of string
theory under consideration, and is given by
\beq
         n ~=~ {f\over 4}\, \alpha' M^2 ~~~~{\rm where}~
           f = \cases{ 1 & for closed strings\cr
                       4 & for open strings.\cr}~
\label{ntoM}
\eeq
Here $\alpha'\equiv M_{\rm string}^{-2}$ is the Regge slope, and the
factor $f$ reflects the different conventional normalizations for
the lengths of closed versus open strings.
The second step is to extract a density $\rho(M)$ from the level
degeneracies $g_n$.
By equating the discrete partition function
          $ Z\equiv \sum_n g_n e^{-\beta M}$
with the continuous partition function
          $ Z\sim \int dM \rho(M) e^{-\beta M}$
in the limit $T\to T_H$, we obtain
\beq
              \rho(M) ~=~ (\half f \alpha' M)\, g_n~.
\eeq
Note, in particular, that we do not divide by $M$ to obtain the density;
rather we multiply by $M$ and adjust the units via $\alpha'$.
Thus, putting the pieces together, we are able to relate the coefficients
$(B,C)$ in (\ref{stringasymform}) to the coefficients $(b,c)$ in
(\ref{asymdensity}),
yielding
\beq
           b= 2B-1~,~~~~~~~~~ c= \half \sqrt{f\alpha'}\, C~.
\eeq
The string Hagedorn temperature is therefore given by
\beq
           \sqrt{\alpha'}\, T_H ~=~ {2\over\sqrt{f}}\, C^{-1}
\eeq
and we find
\beqn
           B ~\leq~ (D+5)/4 ~~~&\Longrightarrow&~~~ \hbox{limiting temperature}
\nonumber\\
           B ~>~ (D+5)/4    ~~~&\Longrightarrow&~~~ \hbox{phase transition}.
\eeqn

The final step is to calculate the coefficients $B$ and $C$ for the different
string theories, taking proper account of the Kaluza-Klein and winding modes
as well as the usual string oscillator modes.
In the case of closed strings,
both $B$ and $C$ receive separate contributions from the
left- and right-moving components of the worldsheet theory;
these contributions are then added together.
In the case of open strings, by contrast, the left- and right-moving
oscillations
are required to conspire to form standing waves, and hence only one such
component (left or right) is sufficient to describe the states of the string.
In either case,
it turns out~\cite{Hardy} that these left-
and right-moving contributions depend
on only the light-cone central charge
$\gamma$ of the appropriate worldsheet conformal field theory and
the modular weight $k$ of its characters:
\beq
        C_{L,R} ~=~ \sqrt{ {2\gamma\over 3} } \,\pi~,~~~~~~
        B_{L,R} ~=~ {3\over 4} - {k\over 2}~.
\eeq
Note that the role of the Kaluza-Klein states
is to leave the light-cone central charge of the conformal field
theory unaffected (thereby preserving the value of $C$), but
to modify the modular weight of the characters (thereby affecting $B$).
This occurs because the summation over Kaluza-Klein modes introduces additional
$\vartheta$-functions,
each with modular weight $+1/2$, into the full string one-loop partition
function.
Thus, while the value of $C$ is unaffected by the compactification
of a given string theory from $D=10$ to $D<10$,
the value of $B$ is changed in a dimension-dependent manner.
It is this effect which was not incorporated into several prior
analyses~\cite{tye,alvarez}.

Given this understanding, we shall now simply quote the results.
For a Type~II string compactified to $D$ spacetime dimensions, we find
\beq
          C~=~ 4\sqrt{2}\pi~,~~~~~
          B~=~ {11\over 2} - {10-D\over 2}~=~{1\over 2}\,(D+1)~.
\label{TypeIIresult}
\eeq
Here the first contribution to $B$ comes from the string oscillator
modes, while the second comes from the Kaluza-Klein modes.
This combined value of $B$ implies a phase transition for
all spacetime dimensions $D\geq 4$
at the temperature $\sqrt{\alpha'} T_H= (2\sqrt{2}\pi)^{-1}$, or
$ T_H \approx  M_{\rm string}/9$.
Likewise, for a heterotic string compactified to $D$ spacetime
dimensions, we find
\beq
          C~=~ 2(2+\sqrt{2})\pi~,~~~~~ B~=~{1\over 2}\,(D+1)~,
\label{heteroticresult}
\eeq
again implying a phase transition at all spacetime dimensions $D\geq 4$
at the slightly lower temperature $\sqrt{\alpha'} T_H =
[(2+\sqrt{2})\pi]^{-1}$,
or $ T_H \approx M_{\rm string}/11$.
In general, the properties of such a phase transition and the physics
beyond the Hagedorn temperature are not well-understood.  Various
discussions can be found in
Refs.~\cite{bowick,tye,alvarez,sundborg,kogan,atickwitten,osorio,tan,others,giddings,modern}.

However, the situation is completely different for Type~I strings.  If we first
consider the contributions from the perturbative open-string sectors
(corresponding to open strings stretched between the compactified nine-branes),
we find
\beq
          C~=~ 2\sqrt{2}\pi~,~~~~~ B~=~{1\over 4}\,(D+1)~.
\eeq
This leads to the same Hagedorn temperature as in the Type~II case;
of course, this is to be expected since the Type~II theory is a
subset of the Type~I theory and corresponds to its closed-string sector.
However, because of the different value of $B$,
 {\it we see that the open string theory has a true limiting temperature for
all
 values of $D$.}
Note, in particular, that this result disagrees with that found in Table~2
of Ref.~\cite{alvarez}, where the contributions of the Kaluza-Klein states
were not taken into account.
Thus, we see that within the context of an open string theory, we
face the prospect of a true limiting Hagedorn
temperature {\it for all values of}\/ $D$.
In other words, all energy pumped into the system goes into exciting
high-mass open-string states rather than into increasing the thermal kinetic
energy of the low-mass string states.

One natural question that arises in the case of open strings
is whether this conclusion is affected by the presence of
non-perturbative Dirichlet $p$-branes and their associated
excitations.  After all, it might seem that since the $p$-branes have
different effective dimensionalities which depend on $p$,
they might give rise to non-perturbative
states whose thermodynamical properties depend on $p$ rather than
on the full spacetime dimension $D$.
Ultimately, however, it can be shown that this is not the case.
Mathematically, this can be seen
by analyzing the partition functions
of the corresponding non-perturbative string sectors.
Physically, however, we can easily see that although a given open string
might have its endpoints restricted to a $p$-brane,
the density of states to which it gives rise is determined
by its {\it excitations}\/, \ie, its varied embeddings
into the external $D$-dimensional spacetime.  Thus, the states that
arise from the potentially non-perturbative $(p_1,p_2)$-sectors
of open-string theory will obey the same properties as those
from the perturbative nine-brane/nine-brane sectors discussed above,
irrespective of the values of $(p_1,p_2)$.

Thus, to summarize, we see that the behavior of various thermodynamic
quantities
depends on the effective spacetime dimension in different ways, depending
on whether we are dealing with closed or open strings.  These results are
summarized
in Table~\ref{tableone}.

\begin{table}[ht]
\centerline{\begin{tabular}{||c||c|c|c||c||}
\hline
  ~ & \multicolumn{3}{|c||}{closed strings} & {open strings} \\
\hline
  ~ & $D=4$ & $D=5$ & $D=6$ & all $D$ \\
\hline
\hline
 $F$ &  + & +  & + &  $-1$ \\
 $U,S$ &  + & + &  + &  $-2$ \\
 $c_V$ &  $-1/2$ & 0 & + &  $-3$ \\
\hline
\end{tabular}}
\caption{Divergence behavior of the free energy $F$,
           the internal energy $U$, the entropy $S$, and the specific heat
         $c_V$ as $T\to T_H$, for both closed and open strings,
         as a function of the number $D$ of uncompactified
          spacetime dimensions.
         For each thermodynamic quantity $X$, we have listed the corresponding
         divergence exponent $x$, defined as 
            $X(T)\sim (T_H-T)^x$ as $T\to T_H$.
         Here $x=0$ indicates the logarithmic behavior 
              $X(T)\sim |\log(T-T_H)|$,
         and `$+$' indicates a non-divergent quantity.}
\label{tableone}
\end{table}

These results have important implications for
Type~I string theories.
Recall that Type~I string theories contain both closed- and open-string
sectors.
The closed strings correspond to the gravitational sector
(as well as those gauge symmetries resulting from compactification
of the higher-dimensional gravity theory).  By contrast, the open strings
give rise to the gauge symmetries resulting from the nine-branes
(Chan-Paton factors).  Thus, we see that within the context of
open-string theories, {\it it is possible for the gravitational
and gauge sectors to experience different thermodynamic behaviors}
as the Hagedorn temperature is approached.
Specifically, we see that it is possible for the gravitational sector
to undergo a Hagedorn phase transition and enter an (unknown) post-Hagedorn
phase, while the Chan-Paton gauge sector instead feels a limiting temperature
with divergent thermodynamic quantities.

This situation might have various cosmological consequences.
For instance, in the pre-big-bang cosmology~\cite{veneziano},
a period of dilaton-driven inflation is ended when the curvature
becomes of order of the string scale, thus preventing the scale
factor of the three-dimensional universe from reaching the singularity.
A smooth transition to the standard hot big-bang cosmology is
supposed to follow. It is possible, though, that in the phase of
high curvature the Kaluza-Klein modes and the oscillator and
winding modes of the string are
efficiently excited.  If these modes thermalize, one might expect that the
resulting temperature is of the order of the string scale,
leading to a Hagedorn phase transition in the gravitational sector.
This stage might change the estimate of the total energy stored
in the quantum fluctuations amplified by the pre-big-bang backgrounds,
which might in turn change the way the universe enters
the radiation-dominated phase.

There are also several novel features in the case of Type~I strings
with large-radius compactifications.
Ordinarily, in a string theory whose compactification radii are close
to the string scale, it is not possible to change the spacetime dimensionality
as a function of the energy scale when the energy is below the
Hagedorn temperature.
This is because, as we have seen, the Hagedorn temperature is
typically an order of magnitude below the string scale.  Thus, the
effective value of $D$ is fixed in such theories.
However, if the string theory in question has an intermediate-scale radius
whose energy scale $R^{-1}$ is substantially below the corresponding string
scale, it is possible, upon increasing the energy and temperature of the
system, to cross the radius threshold and thereby effectively increase
the value of $D$.

This raises some intriguing possibilities.  The electroweak
phase transition might have taken place when the temperature of the universe
was not far below the Hagedorn temperature.  Furthermore,
we see from Table~\ref{tableone}
that our findings can have an important effect on the behavior of the specific
heat $c_V$ as the temperature of the system is increased.
Specifically, we can easily imagine a situation in which the specific
heat is driven towards large values as energy is pumped into the system,
until the energy exceeds the radius threshold and a new dimension opens up.
This in turn could change the thermodynamics of the system in such a way
that large amounts of entropy are
suddenly ``released'' when the universe cools
and the number of extra dimensions decreases.   This in turn could dilute
the densities of unwanted relics such as domain walls and magnetic monopoles
which
were created at earlier epochs.  Indeed, this is the stringy analogue of the
idea of
large entropy generation via dimensional compactification in field
theory~\cite{rocky}.  Moreover, in the present case,
the decay of an exponentially large number of massive string states
may be of further help.

Finally, note that in the case of open strings, the Hagedorn
phenomenon provides us with a natural way of generating extremely large
values of thermodynamic quantities such as energy and entropy --- indeed,
for temperatures approaching the Hagedorn temperature, these values will
be much larger than would have been expected without string theory.
This simple fact may also have important cosmological implications.
For example, in order to solve the smoothness and flatness problems of the
standard big-bang cosmology~\cite{book}, one requires that the patch
containing our present observed universe contain an entropy greater
than about $10^{88}$.
Therefore, creating a large amount of entropy close to the Hagedorn
temperature may help in explaining why the universe looks so smooth and
flat to us.

However, as we shall now discuss,
perhaps the most useful implication of this fact is that
it may be used to generate and stabilize a large radius of compactification.

\subsection{Thermal generation of a large compactification radius}

Let us now turn our attention to perhaps the most important problem that
affects any discussion of theories with large extra spacetime dimensions:
the generation (and indeed the stabilization) of such a large radius of
compactification.
While there are many mechanisms that might be imagined (see, for example,
Refs.~\cite{BV,Sundrum}),
in this section we shall explore a relatively simple idea.
In string theory, spacetime supersymmetry forces all compactification radii
to act as moduli --- \ie, they have an exactly flat potential to all orders
in perturbation theory.  However, because bosons and fermions feel different
statistics at finite temperatures, thermal effects necessarily
break supersymmetry.  It is therefore natural to expect that thermal
effects can produce a potential for the radii of compactification.

In this section, we shall explore this idea within the context of a simple
toy string model, namely the ten-dimensional supersymmetric Type~I $SO(32)$
string evaluated at finite temperature and toroidally compactified down to nine
dimensions with a single radius of compactification.
As a function of the temperature, we shall calculate the one-loop potential
for the radius.
It turns out that the behavior we shall find is generic, even for
compactifications down to four dimensions.

Unlike the case in field theory,
three new features arise in the case of Type~I string theory.
The first, of course, is the Hagedorn phenomenon, discussed above.
The second is the presence of not only momentum modes whose energies are
inversely proportional to the compactification radius $R$,
but also {\it winding}\/ modes
whose energies grow linearly with $R$.  Such winding modes arise
in only the closed-string (gravitational) sector of the Type~I theory.
Finally, in Type~I string theory, the usual one-loop field-theory diagram
now generalizes to receive four separate contributions from the four possible
unoriented one-loop topologies:  the torus, the Klein bottle, the cylinder,
and the M\"obius strip.  Each gives rise to a different radius-dependence.

In this paper, we shall not review the construction of
the $SO(32)$ Type~I theory or the derivation of these one-loop amplitudes.
Instead, we shall merely write down the results.  In order to do this,
we first define the circle-compactified  partition function
\beq
      Z_R(\tau) ~\equiv~ \sum_{m,n} \,{1\over \overline{\eta}\eta} \,
                \overline{q}^{(ma-n/a)^2/4}  \,q^{(ma+n/a)^2/4}
\eeq
where $q\equiv e^{2\pi i \tau}$,
where
\beq
        \eta(\tau) ~\equiv~ q^{1/24} \,\prod_{n=1}^\infty (1-q^n)~,
\eeq
and where we have defined the inverse dimensionless radius
$a\equiv \sqrt{\alpha'}/R$.
Here the sums over $(m,n)$ respectively represent the contributions
from Kaluza-Klein momentum and winding modes of the string.
Next, we define four related circle-compactified functions which
are equivalent to $Z_R(\tau)$ except for various restrictions on their
summation variables:
\beqn
      {\cal E}'_0 &\equiv& \lbrace   m\in\IZ, ~n ~{\rm even} \rbrace
\nonumber\\
  {\cal E}'_{1/2} &\equiv& \lbrace   m\in\IZ+\half, ~n ~{\rm even} \rbrace
\nonumber\\
      {\cal O}'_0 &\equiv& \lbrace   m\in\IZ, ~n ~{\rm odd} \rbrace \nonumber\\
  {\cal O}'_{1/2} &\equiv& \lbrace   m\in\IZ+\half, ~n ~{\rm odd} \rbrace ~.
\eeqn
Third, we define the so-called characters of the transverse Lorentz
group $SO(8)$ in terms of the Jacobi $\vartheta$-functions given in
(\ref{thetadefs}):
\beqn
 \chi_I &=&  \half\,(\tthree^4 + \tfour^4)/\eta^4 \nonumber\\
 \chi_V &=&  \half\,(\tthree^4 - \tfour^4)/\eta^4 \nonumber\\
 \chi_S &=&  \half\,(\ttwo^4 + {\vartheta_1}^4)/\eta^4 \nonumber\\
 \chi_C &=&  \half\,(\ttwo^4 - {\vartheta_1}^4)/\eta^4 ~.
\eeqn
These characters correspond to spacetime scalars, vectors, spinors,
and conjugate spinors of $SO(8)$ respectively.
Note that $\chi_V=\chi_S=\chi_C$ as
functions of $\tau$;  this is a manifestation of $SO(8)$ triality.
Finally,
we note that the incorporation of finite-temperature effects can be achieved
in the usual way, via the introduction of Matsubara modes which can be
viewed as the Kaluza-Klein modes corresponding to the compactification
of an additional (time) direction on a circle
of radius $R_T=\beta/2\pi = (2\pi T)^{-1}$.
We will therefore also define the corresponding inverse-radius variable
\beq
      a_T ~\equiv~ {\sqrt{\alpha'} \over R_T} ~=~ 2\pi \,{T\over M_{\rm
string}}~
\label{aTdef}
\eeq
and use the same circle-compactified functions as before.
Note from (\ref{aTdef}) that $2\pi T$ serves as a ``thermal mass''.
Also note, in this regard, that the Hagedorn temperature for the Type~I system
corresponds to $a_T^\ast= 1/\sqrt{2}$.  We shall
therefore be forced to restrict our attention to values
$a_T < 1/\sqrt{2}$ in what follows.

Given these definitions, it is then straightforward to write down
the string contributions to the one-loop effective potential $V$
(or equivalently, to the so-called cosmological constant $\Lambda\equiv V/T$)
arising from the four sectors
corresponding to the torus, the Klein-bottle, the cylinder, and the
M\"obius-strip one-loop amplitudes.
We find
\beq
          \Lambda ~=~ \Lambda_T + \Lambda_K + \Lambda_C + \Lambda_M
\eeq
where these individual contributions are given as~\cite{ADS,Julie}:
\beqn
    \Lambda_T &=& -\half \int_{\cal F} {d^2\tau\over ({\rm Im}\,\tau)^5}
         \, (\overline{\eta}\eta)^{-6} \, \biggl\lbrace
      \lbrack \chibar_V\chi_V + \chibar_S\chi_S \rbrack \, \calE'_0(a_T)
       ~ +~
         \lbrack \chibar_I\chi_I + \chibar_C\chi_C \rbrack
\,\calO'_{0}(a_T)\nonumber\\
      &&~~~~~~-~ \lbrack \chibar_V\chi_S + \chibar_S\chi_V \rbrack
\,\calE'_{1/2}(a_T)
       ~-~ \lbrack \chibar_I\chi_C + \chibar_C\chi_I \rbrack
           \,\calO'_{1/2}(a_T) \biggr\rbrace~ Z_R(\tau)\nonumber\\
   \Lambda_K &=& -\half \int_0^\infty  {dt\over t^5} \,
          \eta(q^4)^{-8}\,(\chi_V-\chi_S)(q^4)\,
              \left(\sum_{m=-\infty}^\infty  q^{2 m^2 a^2_T} \right)
              \left(\sum_{m=-\infty}^\infty  q^{2 m^2 a^2} \right)\nonumber\\
   \Lambda_C &=& -\half N^2\,\int_0^\infty  {dt\over t^5} \,
          \eta(q)^{-8}\,
           \biggl\lbrace \chi_V(q)\, \sum_{m=-\infty}^\infty q^{m^2 a^2_T}
      ~-~\chi_S(q)\, \sum_{m=-\infty}^\infty q^{(m+1/2)^2 a^2_T} \biggr\rbrace
\nonumber\\
            && ~~~~~~~\times\, \sum_{m=-\infty}^\infty q^{m^2 a^2}\nonumber\\
   \Lambda_M &=& \half N\,\int_0^\infty  {dt\over t^5} \,
          \eta(-q)^{-8}\,
           \biggl\lbrace \chi_V(-q)\, \sum_{m=-\infty}^\infty q^{m^2 a^2_T}
             ~-~\chi_S(-q)\, \sum_{m=-\infty}^\infty q^{(m+1/2)^2 a^2_T}
                      \biggr\rbrace \nonumber\\
            && ~~~~~~~\times\, \sum_{m=-\infty}^\infty q^{m^2 a^2}~.
\label{LambdaTKCM}
\eeqn
In each case, these (dimensionless) cosmological constants are given in
units of $\half {\cal M}^8$, where ${\cal M}$ is the reduced string scale
$M_{\rm string}/2\pi$;  thus they have the dimensions of inverse eight-volumes
in our nine-dimensional theory.
In (\ref{LambdaTKCM}),
$N=32$ is the number of nine-branes in the theory, 
with $q\equiv e^{2\pi i\tau}$
for the torus amplitude and $q\equiv e^{-\pi t}$ for the Klein-bottle,
cylinder, and
M\"obius-strip amplitudes.
Note that while the torus amplitude receives contributions from both
Kaluza-Klein
momentum and winding modes,
the string orientifold projection removes all winding modes, and consequently
the remaining three amplitudes receive contributions from only the
compactification
momentum modes.  Also note that $\Lambda_K=0$ as a result of
the identity $\chi_V=\chi_S$, signalling that spacetime supersymmetry is
not broken by thermal effects in the
Klein-bottle sector of the theory.  However, in each of the remaining sectors,
the
supersymmetry is broken through the relative half-shift in the Matsubara
frequencies
between bosonic states (corresponding to $\chi_{I,V}$) and fermionic states
(corresponding to $\chi_{S,C}$).

Note that (\ref{LambdaTKCM})
is simply the stringy generalization of
field-theoretic expressions such as (\ref{pot1}).
This can be shown (for example, for the cylinder amplitude) as follows.
For simplicity, we shall take $R\to \infty$ (or $a\to 0$).
Since $\chi_V=\chi_S$, we can write the resulting
cylinder amplitude as 
\beq
        \Lambda_C ~\sim~ T^9 \,\int_0^\infty {ds\over s^{11/2}}\,
       \left({\vartheta_2^4 \over \eta^{12}}\right)(2\pi^2 is / a_T^2)\,
      \left\lbrack \vartheta_3(4\pi^2 i s)-\vartheta_2(4\pi^2 is)\right\rbrack~
\label{aneq}
\eeq
where $s\equiv a_T^2 t/(4\pi)$.
In the field-theory limit $T\ll M_{\rm string}$ (or $a_T\ll 2\pi$),
the contribution $\vartheta_2^4/ \eta^{12}$ from the string oscillators
goes to a constant. 
By using the identity
\beqn
           && \int_0^\infty {d^{D-1}k \over (2\pi)^{D-1}} \ln\left(
          {1-e^{-\beta E}\over 1+e^{-\beta E}}\right) ~=~ \nonumber\\
           &&~~~~~- {1\over 2 (4\pi)^{(D-1)/2}}\,\int_0^\infty 
            {ds\over s^{(D+1)/2}} \, e^{-M^2 s}
      \left\lbrack \vartheta_3(4\pi is T^2) - 
            \vartheta_2(4\pi is T^2)\right\rbrack~,
\eeqn
we see that (\ref{aneq}) yields the effective potential of a 
supersymmetric ten-dimensional
field theory at finite temperature.  Thus, we see that the field-theory 
limit of our string calculation corresponds to $T\ll M_{\rm string}$.

At a formal level, it is interesting to note that the $R\to \infty$ limit of
this model is equivalent to the nine-dimensional non-supersymmetric
Type~I $SO(32)$ ``interpolating model'' which was constructed in
Ref.~\cite{Julie}
and used in order to construct strong/weak coupling Type~I duals for
non-supersymmetric heterotic strings.  Indeed,
as shown in Ref.~\cite{ADS}, the temperature $T$ here plays
the role of the compactification radius $R$ of Ref.~\cite{Julie}
via the relation $T= 1/\pi R$.
However, unlike Ref.~\cite{Julie}, we shall hold $T$ fixed and
calculate the dependence of $\Lambda$ on $R$ for fixed $T$.
This will ultimately yield the desired finite-temperature potential for the
radius $R$.

Evaluating these integrals is relatively straightforward,
and proceeds using methods analogous to those discussed in Ref.~\cite{Julie}.
The results are shown in Figs.~\ref{interpfigone} and \ref{interpfigtwo}.
In Fig.~\ref{interpfigone}, we show the separate torus, cylinder, and
M\"obius-strip contributions to the effective potential, as well as their
total.  For this calculation we have taken $a_T=1/2$.
It is clear that the cylinder contribution dominates the sum, due to the
large multiplicity of nine-branes in the theory.
It is also clear that while the torus contribution exhibits the expected
$T$ duality under which $a\to 1/a$ or $R\to \alpha'/R$ (due to the presence
of both momentum and winding modes in this sector),
the open-string contributions do not.
Instead, as $R$ becomes small, the compactified momentum modes
become extremely heavy and their contributions to the potential vanish.  Thus,
as $R\to 0$, all radius dependence essentially ``freezes out'' of these
open-string
contributions, and their contributions to the total effective potential become
flat in this regime.

It is important to note that although we are plotting values of $R$
for $R/\sqrt{\alpha'} <1$ as
well as $R/\sqrt{\alpha'}>1$, their physical interpretation is completely
different.
For $R/\sqrt{\alpha'}>1$, the radius is larger than the fundamental string
length,
so the proper interpretation is indeed that of a compactified Type~I theory.
For $R/\sqrt{\alpha'}<1$, by contrast, the radius is smaller than the
fundamental
string length.  In this case the proper interpretation
is the $T$-dual one, namely that of a compactified Type~I$'$ theory where the
radius
$R$ now signifies the distance perpendicular to the nine-branes.
Thus, in this way our analysis is applicable to both ``universal'' extra
dimensions as well as those felt only by gravity.

\begin{figure}
\centerline{ \epsfxsize 3.25 truein \epsfbox {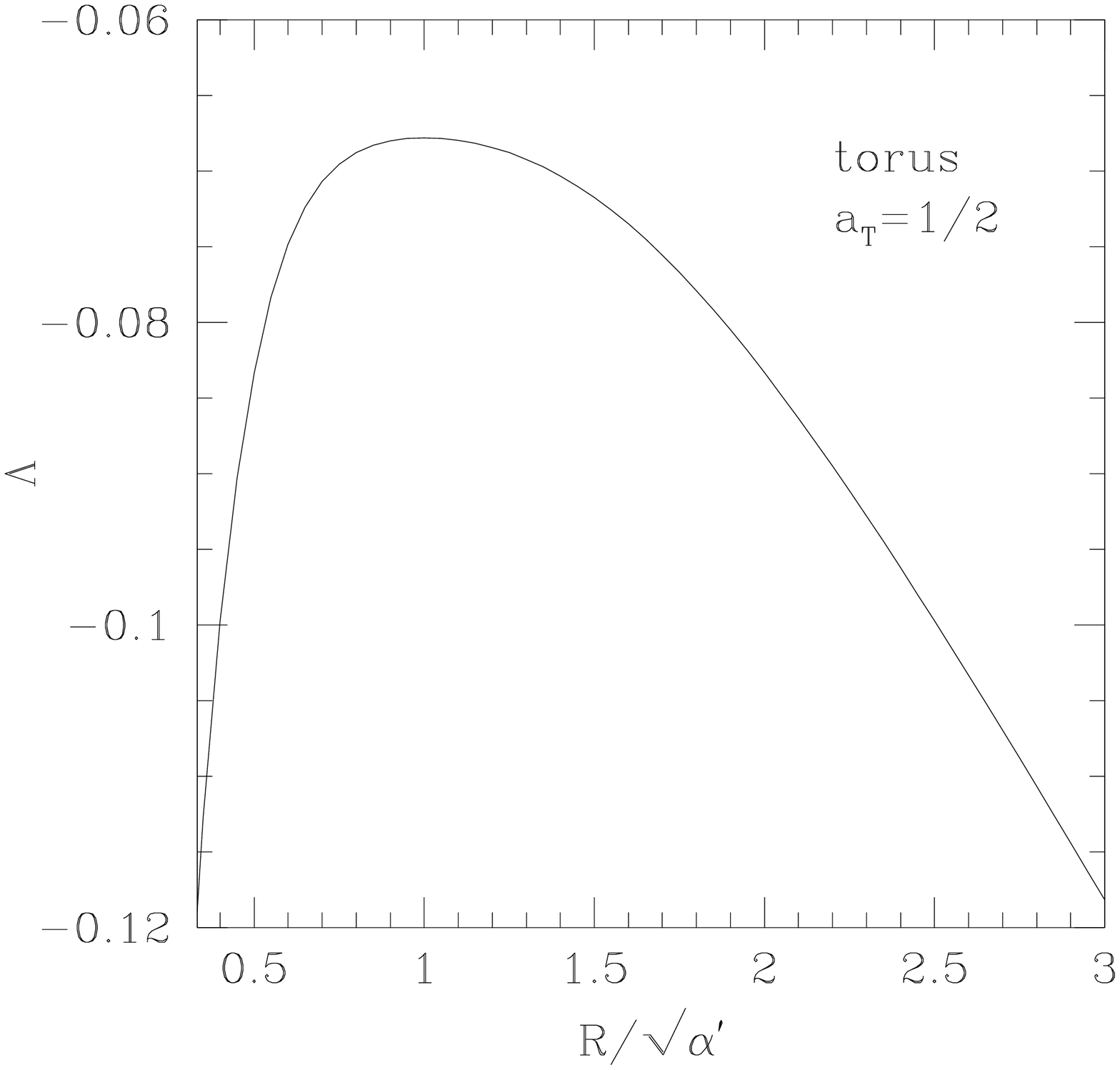}
             \epsfxsize 3.25 truein \epsfbox {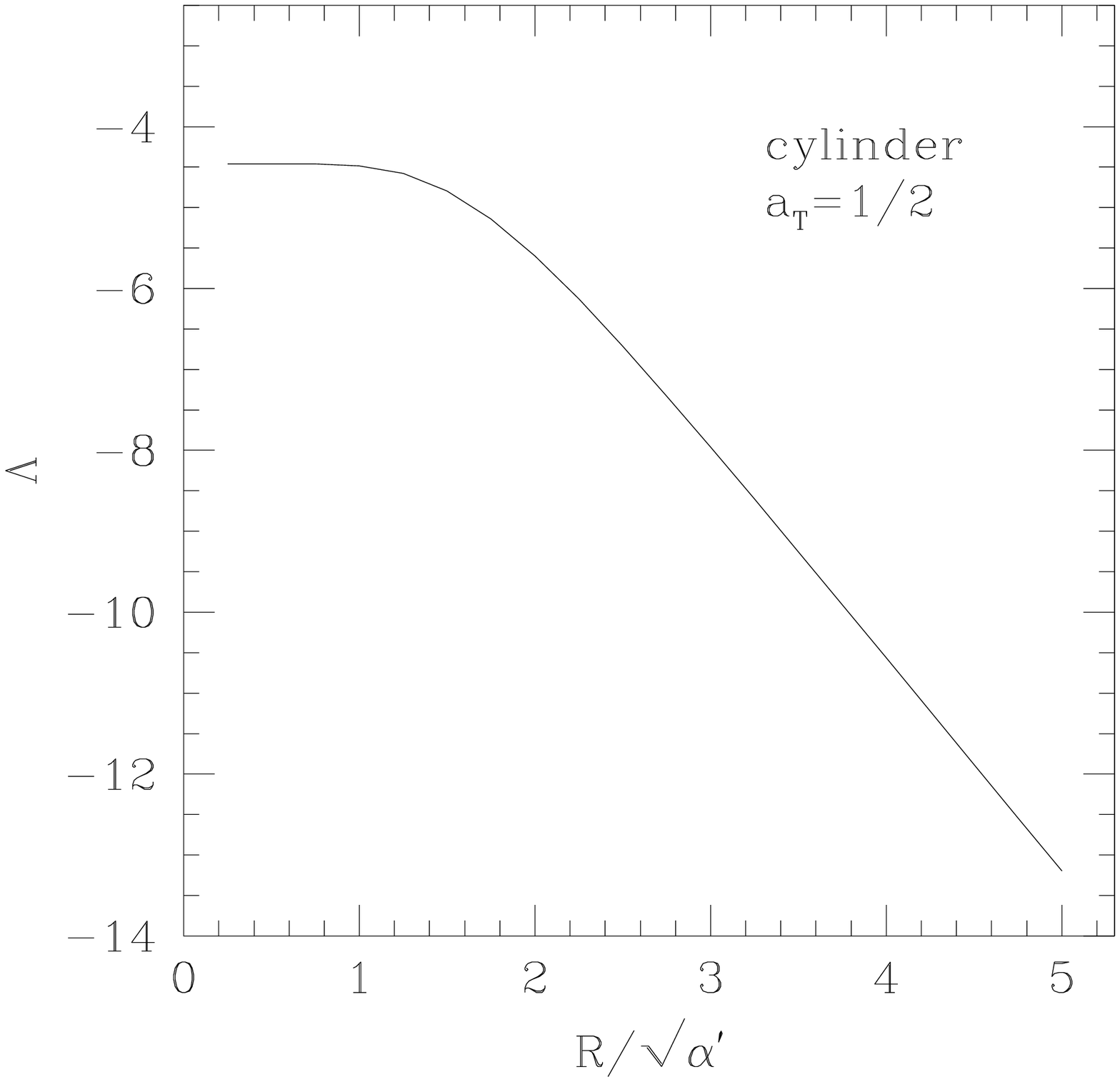}}
\centerline{ \epsfxsize 3.25 truein \epsfbox {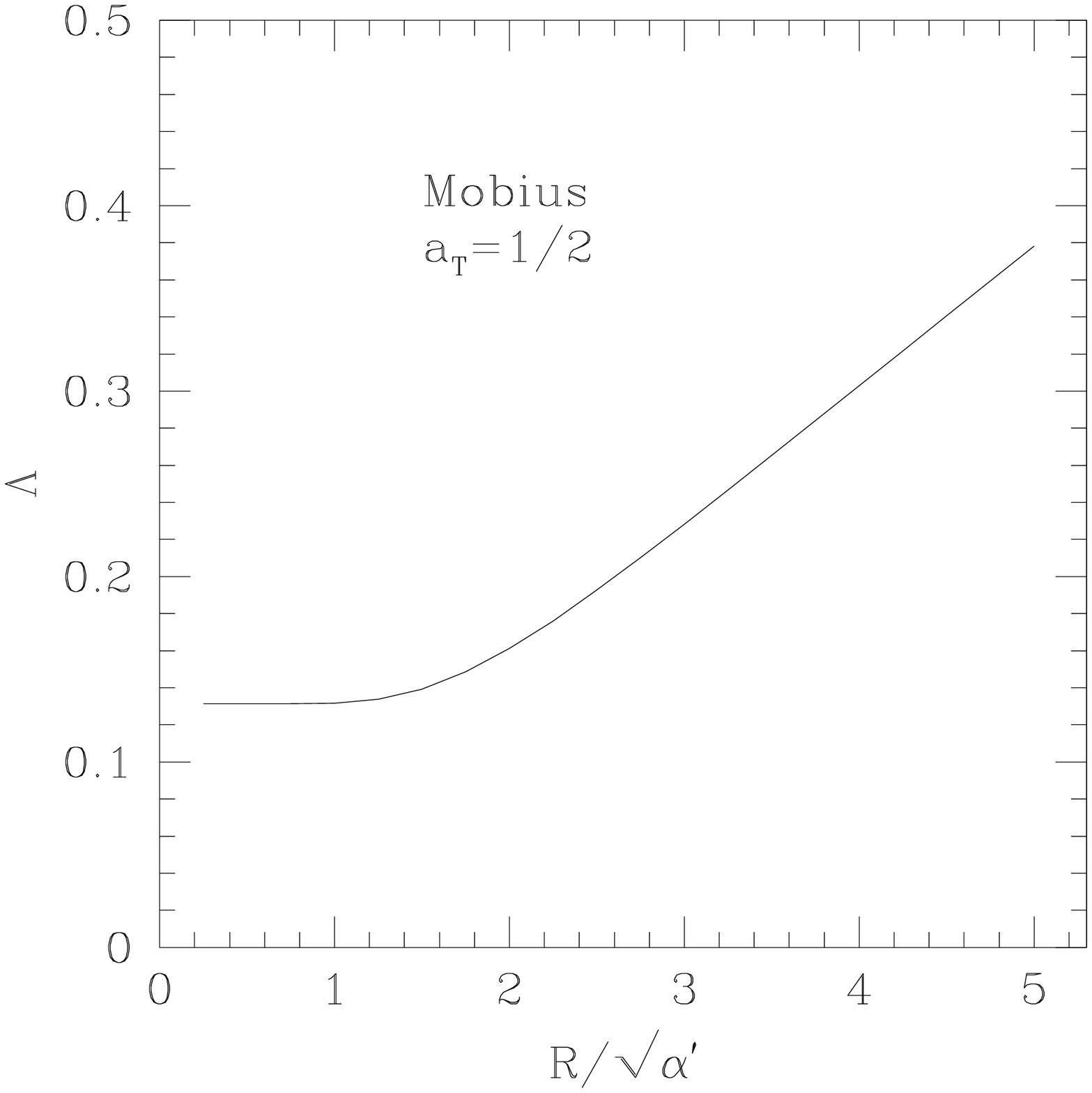}
             \epsfxsize 3.25 truein \epsfbox {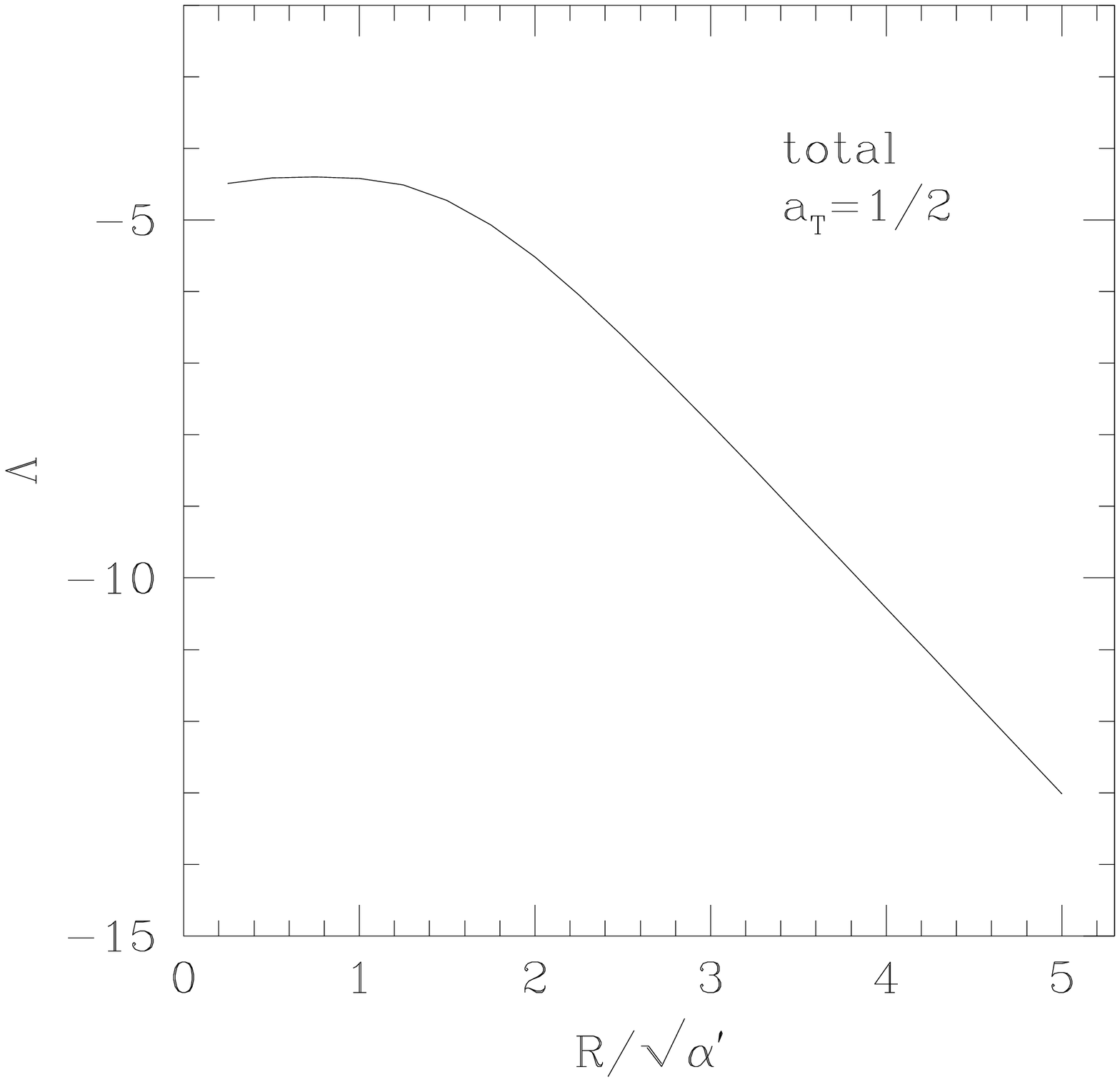}}
\caption{Individual contributions to the effective potential
     as a function of the radius of compactification,
     evaluated at the temperature
     $a_T=1/2$ or $T/M_{\rm string}=1/4\pi$.
     The total effective potential (lower right) shows a clear
      tendency to push the radius out to large values.}
\label{interpfigone}
\end{figure}

\begin{figure}[ht]
\centerline{ \epsfxsize 4.0 truein \epsfbox {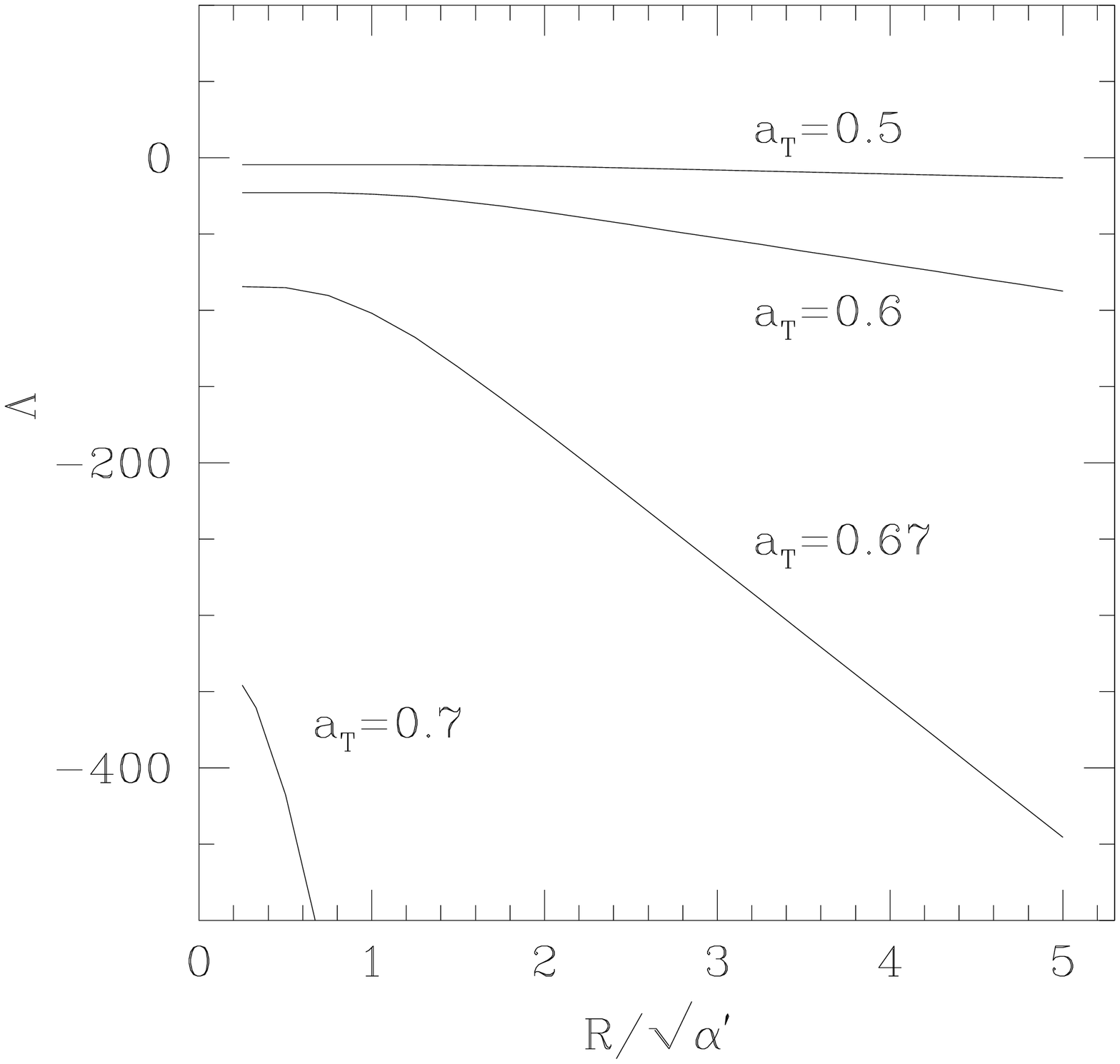}}
\caption{The effective potential as a function of the radius
      of compactification, for temperatures ranging from
      $a_T=0.5$ (or $T/M_{\rm string}=1/4\pi$) to
      $a_T=0.7$ (or $T/M_{\rm string}=7/20\pi$).
      In all cases, the radius is pushed out to large values.
      Note that the potential diverges to negative values at the
       Hagedorn temperature $a_T=1/\sqrt{2}\approx 0.707$.}
\label{interpfigtwo}
\end{figure}

All of the plots in Fig.~\ref{interpfigone} are calculated for $a_T=1/2$.
In Fig.~\ref{interpfigtwo}, by contrast, we show how the total effective
potential depends on the temperature $a_T=2\pi T/M_{\rm string}$.
For small temperatures, we see that the effective potential becomes flat,
reflecting the restoration of spacetime supersymmetry in this limit.
For larger temperatures, however, we see that the effective potential
becomes quite strong and steep.   Of course, as expected, this
potential ultimately diverges towards negative infinity for all radii $R$ as
$a_T\to a_T^\ast=1/\sqrt{2}\approx 0.707$.  This is of course simply
the Hagedorn temperature of the theory, and reflects the contribution
of a tachyonic Matsubara winding-mode state
which appears in the torus amplitude at this
temperature~\cite{kogan,atickwitten}.

It is important to interpret the results in Figs.~\ref{interpfigone}
and \ref{interpfigtwo} correctly.
It is clear, of course, that in all cases,
 {\it the finite-temperature effects set
up an instability which pushes the radius away from the fundamental
string scale at $R=\sqrt{\alpha'}$ and towards either extremely large
or extremely small values}\/.  (In this connection, recall that
all of these potentials go to $-\infty$ as $R\to 0$.)
Thus, we see that finite-temperature effects can provide a natural
mechanism for generating hierarchically large or small radii within the context
of Type~I string theory.
Moreover,  we also see that finite-temperature effects tend to
render the self-dual point $R=\sqrt{\alpha'}$
unstable.  Our findings might therefore have important implications
for the pre-big-bang scenario where the sizes of the extra dimensions
are usually thought to converge towards their respective self-dual values.

Of course, there are several important caveats that must be
mentioned if one tries to implement this mechanism within a cosmological
model.  First of all, these potentials represent only the contributions that
come
from thermal effects.  While these might be supposed to dominate at high
temperatures (\eg, near the Hagedorn temperature), there will be other
effects at lower temperatures that will come into play, such as the
zero-temperature potentials that arise due to
ordinary supersymmetry-breaking effects.
Furthermore, as we shall discuss below,
entropy conservation can be expected to play an important
role if the universe undergoes an adiabatic evolution.
However, it is interesting that these finite-temperature effects
by themselves are capable of generating either large or small radii at
an early epoch, while the universe is presumably still dominated by thermal
behavior.  This mechanism could therefore be useful
in setting up the initial large- or small-radius {\it pre-conditions}\/
before other potentials come into play.

Although these potentials clearly
generate large radii of compactification, this
still leaves one further question unanswered:
how are such radii ultimately {\it stabilized}\/?

While one can imagine many possible effects that could
intercede and stabilize the compactification radii
as the universe cools,
one natural way to stabilize the radii at all temperatures is
through the constraint of entropy conservation.  Indeed, such a constraint
is appropriate when the universe undergoes an adiabatic evolution.
Given the cosmological constants $\Lambda$ that we have calculated,
it is a straightforward procedure to calculate the total entropy $S$
via the relation
\beq
         S ~=~ -{\partial V\over \partial T} ~=~
          - \left( \Lambda + T {\partial\over \partial T}\Lambda\right)~.
\eeq
We then obtain the results shown in Fig.~\ref{entropyfig}(a) through
Fig.~\ref{entropyfig}(e).
Note that $S$ has the same units as $\Lambda$, namely those
of an inverse eight-volume, and should be interpreted as an entropy density
with respect to the eight-dimensional flat space.
However, this quantity represents the {\it total}\/ entropy with respect
to the compactified dimension with radius $R$.

\begin{figure}
\centerline{ \epsfxsize 5.0 truein \epsfbox {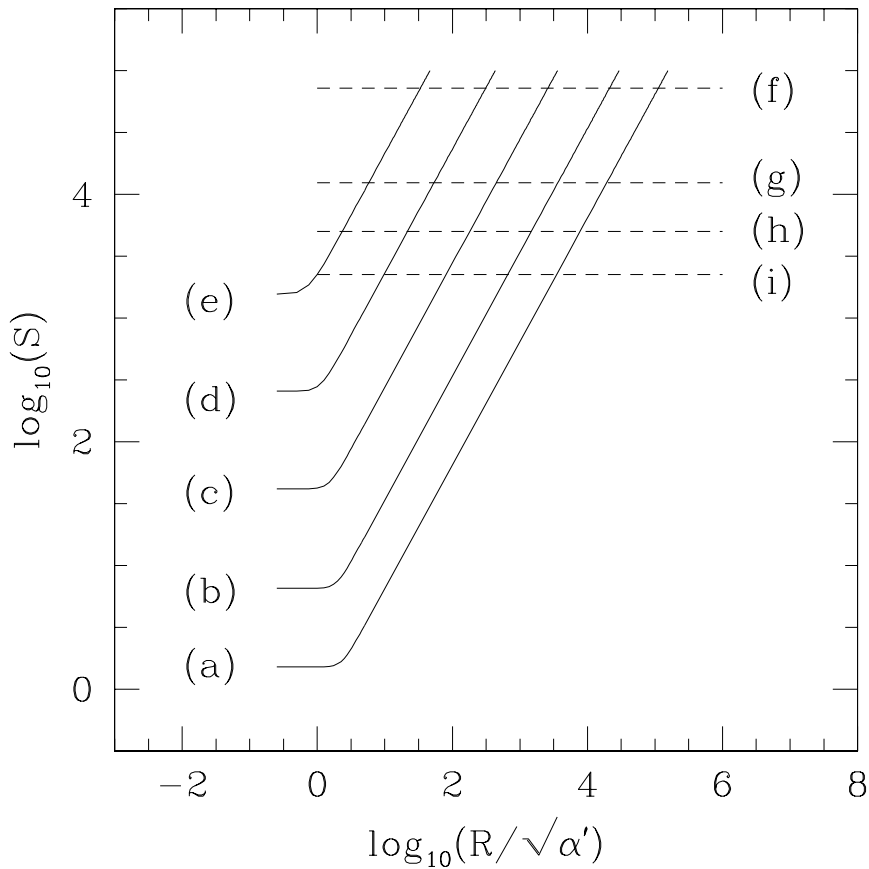}}
\caption{ ~~{\it Solid lines}\/:~Entropy
     as a function of the string compactification radius,
     for temperatures
         (a)~$a_T=0.333$;
         (b)~$a_T=0.4$;
         (c)~$a_T=0.5$;
         (d)~$a_T=0.6$;
         and (e)~$a_T=0.667$.
     {\it Dashed lines}\/:~ Values of the entropy
         for temperatures (f)~$a_T=0.7$; (g)~$a_T=0.69$;
         (h)~$a_T=0.68$; and (i)~$a_T=0.667$, all
         calculated at $R/\sqrt{\alpha'}=1$ and held
         constant as a function of radius.
      If we assume that a cooling phase of the universe
       becomes adiabatic (entropy-conserving) at a given
     initial temperature [(f) through (i)] when the
          compactification radius is at the string scale,
         then hierarchically large compactification radii are
       generated at lower temperatures [(a) through (e)].
       Note that it is the Hagedorn phenomenon at
        $a^\ast_T=1/\sqrt{2}\approx 0.707$ that leads to
      the dramatic rise in the initial entropy as a function
     of temperature which
     in turn generates such hierarchically
        large radii of compactification.  }
\label{entropyfig}
\end{figure}

It is clear from these plots that the entropy $S$ rises linearly as a
function of the compactification radius in
the limit $R/\sqrt{\alpha'}\gg 1$.  It also clear that
the entropy rises
dramatically as a function of the temperature.
Of course, this behavior is not surprising,
for we expect on the basis of field theory alone
that in a ten-dimensional spacetime where one
dimension has been compactified (such as in our toy string model),
the total entropy should scale as
\beq
               S ~\sim~ R\, T^{9} ~
\label{FTentropy}
\eeq
for $RT\gg 1$.
However, the important point is that in string theory,
this dependence is even stronger than in field theory.
Indeed, this enhanced dependence becomes increasingly evident
as the temperature approaches the Hagedorn temperature $T_H$,
for in this limit the ``stringy'' behavior begins to dominate
and we expect from the results in Table~\ref{tableone} that
\beq
              S~\approx~ {S_0\over (T_H-T)^2}~~~~~~
                   {\rm as}~~~T\to T_H~
\label{Shag}
\eeq
for some (radius-dependent) $S_0$.
This behavior is shown from Fig.~\ref{secondentropyfig},
where we plot the entropy
as a function of the temperature,
with fixed radius $R/\sqrt{\alpha'}=1$.
For small temperatures
$a_T\lsim 0.5$, the entropy indeed
grows as a power of the temperature, in accordance with
(\ref{FTentropy}).  For larger temperatures,
however, the entropy begins to exhibit the Hagedorn behavior
(\ref{Shag}), with $S_0\approx 3.63$ and
$T_H=1/\sqrt{2}$.

\begin{figure}[ht]
\centerline{ \epsfxsize 4.0 truein \epsfbox {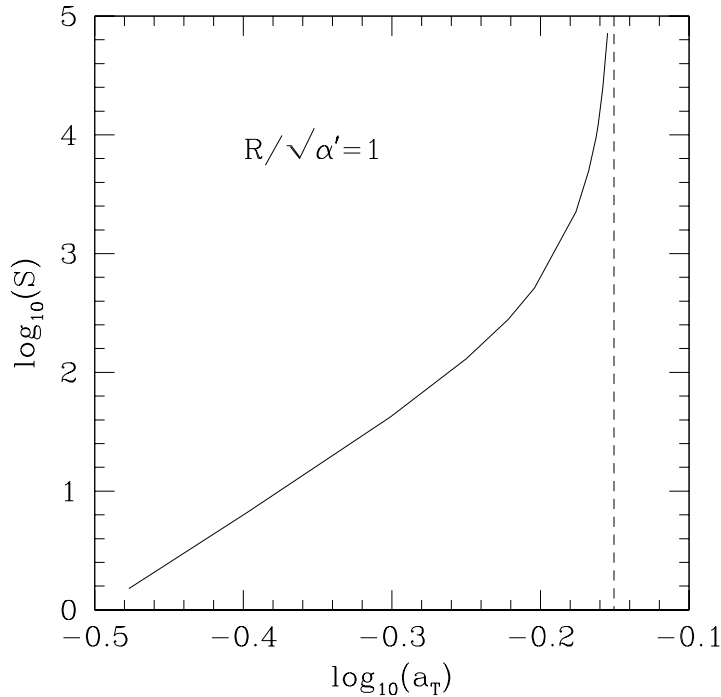}}
\caption{~~{\it Solid line}\/:~ Entropy as a function of the temperature,
evaluated
      for fixed radius $R/\sqrt{\alpha'}=1$.
        For small temperatures, the entropy behaves as expected
         in field theory, while the stringy
        Hagedorn behavior becomes dominant at higher temperatures.
         {\it Dashed line}\/:~  the Hagedorn limiting temperature.}
\label{secondentropyfig}
\end{figure}

Combining these two observations, we see that this leads
to a natural cosmological
mechanism for generating and stabilizing a hierarchically
large radius of compactification.  Of course, we cannot
speculate with any certainty about the nature of the physics at
the Hagedorn temperature.  However, let us imagine the universe
cooling below this temperature.
We shall begin by assuming a radius of compactification near
the string scale, so that $R/\sqrt{\alpha'}=1$.
As the universe cools, we expect that at some temperature $T_{\rm ad}$
the evolution becomes adiabatic, so that the total entropy is conserved.
Presumably this might happen quite early, while the temperature is
still relatively close to the Hagedorn temperature and the entropy
is therefore extremely high.
In Figs.~\ref{entropyfig}(f) through Fig.~\ref{entropyfig}(i),
for example, we have illustrated various values of the (fixed) entropy
that would result.
We then find that large compactification radii are generated
for temperatures below $T_{\rm ad}$.
For example, consulting Fig.~\ref{entropyfig}, we see that
if $T_{\rm ad}$ corresponds to $a_T=0.7$ (as shown in
Fig.~\ref{entropyfig}(f)),
then the compactification radius will have grown to $\gsim 10^5$ in
string units by the time the temperature has dropped to
$a_T=1/3$.
Further cooling will produce compactification radii that are even
hierarchically larger.

Once again, there are several important comments and caveats that
must be mentioned in this connection.
First, it may seem that much of this radius-enhancement effect is
purely field-theoretic.  After all, according
to (\ref{FTentropy}), entropy conservation alone implies the relation
$R_1/R_2=(T_2/T_1)^9$.  However, after compactification to four dimensions
(and assuming that all six compactified dimensions have equal compactification
radii $R$),
the total entropy scales as $S\sim R^6 T^9$, whereupon the field-theoretic
entropy/temperature scaling relation weakens
to $R_1/R_2=(T_2/T_1)^{3/2}$.
By contrast, for sufficiently high temperatures,
the string-theoretic scaling behavior
(\ref{Shag}) is, as we have seen in Sect.~3.1, valid for all spacetime
dimensions
 {\it regardless}\/ of their radii of compactification.
Thus, within the context of string theories near the Hagedorn temperature,
we have a natural mechanism for
boosting the total entropy to values that exceed those
possible in field theory, and this in turn can generate large compactification
radii as the universe cools.

Finally, we must also mention another important caveat.  In the above analysis,
we assumed that our underlying nine- or four-dimensional spacetime
is flat, with fixed (infinite) radius.  Of course, in a realistic cosmological
setting,
this will not be the case, and we can expect to deal with at least {\it two}\/
radii, $R_>$ and $R_<$, corresponding to the three
large spatial dimensions and six small dimensions respectively.
In this case, the field-theoretic entropy relation becomes $S\sim R_>^3 R_<^6
T^9$.
In string theory, of course, the entropy continues to diverge at the Hagedorn
temperature.  Thus, we once again expect to generate hierarchically large
values
of entropy at large temperatures.
However, the implications of this fact
for the generation of large compactification radii
will depend on how this extra entropy is ultimately distributed
between the large and small dimensions.
This in turn will
depend on some additional outside input, such as the dynamics of
the large radii as a function of time or temperature.
For example, it might be that at primordial epochs,
some of the extra spatial dimensions are somehow frozen or contracting very
slowly.  Under these circumstances, if the universe is
cooling, the change of the size of the radius of the remaining extra
dimension(s) may be dramatic near the Hagedorn transition.

However, as pointed out in Ref.~\cite{BV}, this issue may be further
complicated
due to various stringy effects, such as the radius-stabilizing effects of
string
winding modes.  Such string modes can wind around the compactified dimensions,
in the process essentially halting their expansion.
Thus, it is far from clear what generic predictions can be made in such cases,
and we leave this issue for further study.

It is also interesting to note that different effective potentials arise
for different sectors of the theory.   In particular, the gravitational sector
feels only the torus
contribution, while the gauge sector feels the full sum of the 
contributions.  This
means that different radii of compactification  might be generated for the
gravitational and the gauge sectors at finite temperature.
For instance, the gauge sector of the theory might feel extra
dimensions compactified at the scale of $({\rm TeV})^{-1}$, as discussed
in Ref.~\cite{DDG}, while gravity might live in a bulk of dimension
of millimeter-length~\cite{Dim}.
Therefore, the dynamics of dimensional compactification might be
different for the gravitational and gauge sectors.
Indeed, we have already seen in Sect.~3.1 that the gravitational (closed)
string sector experiences only a phase transition at the Hagedorn temperature,
while the gauge (open) string sector actually feels a limiting temperature.
This might have important  implications for some crucial cosmological
issues such as the production of gravitons in the bulk and
the transition to the standard hot big-bang.

Indeed, many other related issues also arise in this context.  For example,
many strong/weak coupling duality relations in string theory provide
non-perturbative connections between open strings and closed strings.
The most famous example of this is the strong/weak coupling duality  
between the $SO(32)$ Type~I (open) string and the $SO(32)$ heterotic
(closed) string.  It would be interesting to understand the implications
of such duality relations for the distinction between 
a Hagedorn limiting temperature and phase transition.
Likewise, it would be interesting to understand how temperature
duality (the symmetry under which $T\to M_{\rm string}^2/T$) emerges
on the heterotic side as the Type~I coupling is increased.
Clearly, as the coupling increases on either side of the duality relation,
interactions (and the effects of non-perturbative $D$-brane states)
should change the thermodynamic behavior.  Of course, it
remains an important issue as to whether duality even holds for $T\not=0$,
since supersymmetry is broken.  Despite certain pieces of evidence 
(see, \eg, Ref.~\cite{Julie}),
it is not yet clear whether duality holds without supersymmetry or
at finite temperature.

Thus, we conclude that within the context of finite-temperature string theory,
thermal effects  provide a natural way of not only generating
 large (in some cases, hierarchically large) radii of compactification, but
also
stabilizing these radii at these large values.  The main new feature relative
to the field-theory case is the generation of large amounts of entropy near
the string Hagedorn temperature.  Of course,
our analysis was only in the context of a nine-dimensional toy string model
in which we took the underlying nine-dimensional spacetime to be fixed
in volume.  However, we expect that the mechanism that we have illustrated
(whereby hierarchically large values of entropy are converted to hierarchically
large compactification radii) is interesting, and may also find application
a more realistic setting.

\bigskip
\medskip
\leftline{\large\bf Acknowledgments}
\medskip

We wish to thank
S.~Dimopoulos, J.~Ellis, J.~March-Russell, M.~Maggiore,
M.~Quir\'os, and G.~Veneziano for useful discussions.


\bigskip
\medskip

\bibliographystyle{unsrt}

\end{document}